\documentclass[twocolumn]{bmcart}% uncomment this for twocolumn layout and comment line below
\usepackage{amsthm,amsmath}
\usepackage[utf8]{inputenc} %unicode support
\usepackage{cite}
\usepackage[pdftex]{graphicx}
\usepackage{amsfonts}
\usepackage{amsmath}
\usepackage{algorithmic}
\usepackage{array}
\usepackage{bbm}

\usepackage{booktabs}
\usepackage{multirow}

\usepackage{siunitx}

\usepackage{pdfpages}

\DeclareMathSymbol{\shortminus}{\mathbin}{AMSa}{"39}

\newcommand\corrige[1]{#1}

\startlocaldefs
\endlocaldefs

\begin{document}

%%% Start of article front matter
\begin{frontmatter}

\begin{fmbox}
\dochead{Research}


\title{Comparison of Semi-supervised Deep Learning Algorithms for Audio Classification}

%%%%%%%%%%%%%%%%%%%%%%%%%%%%%%%%%%%%%%%%%%%%%%
%%                                          %%
%% Enter the authors here                   %%
%%                                          %%
%% Specify information, if available,       %%
%% in the form:                             %%
%%   <key>={<id1>,<id2>}                    %%
%%   <key>=                                 %%
%% Comment or delete the keys which are     %%
%% not used. Repeat \author command as much %%
%% as required.                             %%
%%                                          %%
%%%%%%%%%%%%%%%%%%%%%%%%%%%%%%%%%%%%%%%%%%%%%%

\author[
  addressref={aff1},                   % id's of addresses, e.g. {aff1,aff2}
%   corref={aff1},                       % id of corresponding address, if any
% noteref={n1},                        % id's of article notes, if any
  email={leo.cances@irit.fr}   % email address
]{\inits{L.C.}\fnm{L\'eo} \snm{Cances}}
\author[
  addressref={aff1},                   % id's of addresses, e.g. {aff1,aff2}
%   corref={aff1},                       % id of corresponding address, if any
% noteref={n1},                        % id's of article notes, if any
  email={etienne.labbe@irit.fr}   % email address
]{\inits{E.L.}\fnm{Etienne} \snm{Labb\'e}}
\author[
  addressref={aff1,aff2},
  corref={aff1},                       % id of corresponding address, if any
  email={thomas.pellegrini@irit.fr}
]{\inits{T.P.}\fnm{Thomas} \snm{Pellegrini}}

%%%%%%%%%%%%%%%%%%%%%%%%%%%%%%%%%%%%%%%%%%%%%%
%%                                          %%
%% Enter the authors' addresses here        %%
%%                                          %%
%% Repeat \address commands as much as      %%
%% required.                                %%
%%                                          %%
%%%%%%%%%%%%%%%%%%%%%%%%%%%%%%%%%%%%%%%%%%%%%%

% IRIT, Paul Sabatier University, CNRS, Toulouse, France

\address[id=aff1]{%                           % unique id
  % \orgdiv{Department of Computer Science},             % department, if any
  \orgname{IRIT, University of Toulouse, CNRS},          % university, etc
  \city{Toulouse},                              % city
  \cny{France}                                    % country
}
\address[id=aff2]{%
%   \orgdiv{Institute of Biology},
  \orgname{Artificial and Natural Intelligence Toulouse Institute (ANITI)},
  %\street{},
  %\postcode{}
  \city{Toulouse},
  \cny{France}
}

%%%%%%%%%%%%%%%%%%%%%%%%%%%%%%%%%%%%%%%%%%%%%%
%%                                          %%
%% Enter short notes here                   %%
%%                                          %%
%% Short notes will be after addresses      %%
%% on first page.                           %%
%%                                          %%
%%%%%%%%%%%%%%%%%%%%%%%%%%%%%%%%%%%%%%%%%%%%%%

%%\note{Sample of title note}     % note to the article
%\note[id=n1]{Equal contributor} % note, connected to author

\end{fmbox}% comment this for two column layout

%%%%%%%%%%%%%%%%%%%%%%%%%%%%%%%%%%%%%%%%%%%%%%%
%%                                           %%
%% The Abstract begins here                  %%
%%                                           %%
%% Please refer to the Instructions for      %%
%% authors on https://www.biomedcentral.com/ %%
%% and include the section headings          %%
%% accordingly for your article type.        %%
%%                                           %%
%%%%%%%%%%%%%%%%%%%%%%%%%%%%%%%%%%%%%%%%%%%%%%%

\begin{abstractbox}

\begin{abstract} % abstract
In this article, we adapted five recent SSL methods to the task of audio classification. The first two methods, namely Deep Co-Training (DCT) and Mean Teacher (MT), involve two collaborative neural networks. The three other algorithms, called MixMatch (MM), ReMixMatch (RMM), and FixMatch (FM), are single-model methods that rely primarily on data augmentation strategies. Using the Wide-ResNet-28-2 architecture in all our experiments, 10\% of labeled data and the remaining 90\% as unlabeled data for training, we first compare the error rates of the five methods on three standard benchmark audio datasets: Environmental Sound Classification (ESC-10), UrbanSound8K (UBS8K), and Google Speech Commands (GSC).
\corrige{In all but one cases, MM, RMM, and FM outperformed MT and DCT significantly, MM and RMM being the best methods in most experiments. On UBS8K and GSC, MM achieved 18.02\% and 3.25\% error rate (ER) respectively, outperforming models trained with 100\% of the available labeled data, which reached 23.29\% and 4.94\%, respectively. RMM achieved the best results on ESC-10 (12.00\% ER), followed by FM which reached 13.33\%.} Second, we explored adding the mixup augmentation, used in MM and RMM, to DCT, MT and FM. In almost all cases, mixup brought consistent gains. For instance, on GSC, FM reached 4.44\% and 3.31\% ER without and with mixup. \corrige{Our PyTorch code will be made available upon paper acceptance at https://github.com/Labbeti/SSLH.}

% \corrige{The source code of our implementation will be available on Github~\footnote{https://github.com/Labbeti/SSLH}.}
\end{abstract}

%%%%%%%%%%%%%%%%%%%%%%%%%%%%%%%%%%%%%%%%%%%%%%
%%                                          %%
%% The keywords begin here                  %%
%%                                          %%
%% Put each keyword in separate \kwd{}.     %%
%%                                          %%
%%%%%%%%%%%%%%%%%%%%%%%%%%%%%%%%%%%%%%%%%%%%%%


% MSC classifications codes, if any
%\begin{keyword}[class=AMS]
%\kwd[Primary ]{}
%\kwd{}
%\kwd[; secondary ]{}
%\end{keyword}

\end{abstractbox}
%
%\end{fmbox}% uncomment this for two column layout

\end{frontmatter}

%%%%%%%%%%%%%%%%%%%%%%%%%%%%%%%%%%%%%%%%%%%%%%%%
%%                                            %%
%% The Main Body begins here                  %%
%%                                            %%
%% Please refer to the instructions for       %%
%% authors on:                                %%
%% https://www.biomedcentral.com/getpublished %%
%% and include the section headings           %%
%% accordingly for your article type.         %%
%%                                            %%
%% See the Results and Discussion section     %%
%% for details on how to create sub-sections  %%
%%                                            %%
%% use \cite{...} to cite references          %%
%%  \cite{koon} and                           %%
%%  \cite{oreg,khar,zvai,xjon,schn,pond}      %%
%%                                            %%
%%%%%%%%%%%%%%%%%%%%%%%%%%%%%%%%%%%%%%%%%%%%%%%%

%%%%%%%%%%%%%%%%%%%%%%%%% start of article main body
% <put your article body there>

\section{Introduction}
Semi-supervised learning (SSL) aims to reduce the dependency of deep learning systems on labeled data by integrating unlabeled data during the learning phase. It is essential since the conception of a large labeled dataset is expensive, dependent on the task to be learned, and time-consuming. On the contrary, the acquisition of unlabeled data is cheaper and quicker regardless of the task to perform.
Using unlabeled data while maintaining high performance can be done in three different ways: i) \textit{consistency regularization}~\cite{NIPS2016_30ef30b6,iclrLaineA17}, which encourages a model to produce consistent prediction whereas the input is perturbed, ii) \textit{entropy minimization}~\cite{miyato2018virtual,NIPS2004_96f2b50b,lee2013}, which encourages the model to output high confidence predictions on unlabeled files, and iii) \textit{standard regularization} by using weight decay~\cite{loshchilov2019decoupled,zhang2018mechanisms}, mixup~\cite{zhang2018mixup} or adversarial examples~\cite{wiyatno2019adversarial}. The most direct approach for SSL is pseudo-labeling~\cite{lee2013}, but since then, many new and better approaches came out such as Mean Teacher (MT)~\cite{tarvainen2018mean}, Deep Co-Training (DCT)~\cite{qiao2018deep}, MixMatch (MM)~\cite{berthelot2019mixmatch}, ReMixMatch (RMM)~\cite{berthelot2020remixmatch}, and FixMatch (FM)~\cite{sohn2020fixmatch}.% Most of them have been tested on image datasets, and their description can be found in section~\ref{sec:ssl}.
% Most often, these methods are designed and evaluated on object recognition tasks in images. 

In previous work~\cite{cances2021comparison}, we compared MT and DCT for the task of audio tagging (AT), a classification task that consists of automatically assigning an audio event label to an audio recording. Both approaches use two neural networks during training. In the present article, we extend our comparison by adapting to AT the three single-model SSL methods MM, RMM and FM.
One difficulty lies in choosing which audio data augmentation techniques to use, that work for different types of sound events and spoken words~\cite{grollmisch2021improving}. The augmentations used on images for object recognition, such as flips and rotations, are most often not relevant for audio data. We compare the error rates on three audio datasets with different scopes and sizes: i) Environmental Sound Classification 10 (ESC-10)~\cite{esc-dataset}, with audio event categories such as dog barking and helicopter, ii) UrbanSound8k (UBS8K)~\cite{ubs8k-dataset}, more specific to urban noises such as car horns, sirens and street music, and iii) Google Speech Commands v2 (GSC)~\cite{speechcommandsv2}, containing spoken words exclusively.

%and finally Audioset ....
% You can find a more detailed description of these datasets in section~\ref{sec:datasets}

% In a second step,  more precise experiments will be done on the impact of augmentation in SSL methods. These experiments will highlight the difference between the contribution of the SSL approaches and the contribution of augmentation. It will be achieved by comparing each SSL methods to their supervised counterparts trained in the same configuration (training parameters, augmentation set, and number of labeled data used).

In MM and RMM, a successful data augmentation technique called mixup~\cite{zhang2018mixup} is used. It consists of mixing  pairs  of  samples, both the data samples and the labels with a random coefficient. We propose to add mixup to the three other SSL approaches, namely MT, DCT and FM, which do not already use it. The results reported in this article will highlight the positive impact of mixup in almost all our experiments.

The article contributions are mainly two-fold: i) the application and comparison of several recent SSL methods for audio tagging on three different datasets, ii) the modification of these methods with the integration of mixup, which resulted in systematic error rate reductions. We shall see that in most cases, MM outperformed the other methods, closely followed by FixMatch+mixup.

% Using 10\% of the labeled files, the modified methods allowed to achieve better performance than fully supervised settings, without augmentation. According to our results, the method achieving the best performance is MixMatch, closely followed by FixMatch+mixup.

The structure of the paper is as follows. Section~\ref{sec:aug_mixup} describes the augmentations we used and the mixup mechanism at the core of the present work. Section~\ref{sec:ssl} describes the five SSL methods, Section~\ref{sec:exp} presents the experimental settings, and finally, Section~\ref{sec:results} presents and discusses the results.

\section{Related work}

Semi-supervised learning (SSL) is a well-known machine learning setting, for which a lot of research has been conducted, before the rise in popularity of deep learning~\cite{zhu2005semi,chapelle2006continuation}. In this work, we explore recent SSL approaches that were proposed in the framework of deep learning, since we use deep neural networks as state-of-the-art classifiers for audio tagging. These new approaches, as we shall see, were driven by the simplicity of incorporating unsupervised loss terms into the cost functions of neural networks~\cite{van2020survey}.

\corrige{\subsection*{Semi-supervised deep learning taxonomy}}
In their SSL survey~\cite{van2020survey}, Van Engelen and colleagues proposed a detailed taxonomy for SSL methods in the framework in deep learning. The algorithms explored in the present article fit in the \textit{intrinsically semi-supervised inductive methods} category, meaning methods that attempt to construct a classifier by directly optimizing an objective function for labeled and unlabeled samples. Most semi-supervised neural networks make use of perturbation-based learning methods, where the training data samples (labeled or unlabeled or both) are perturbed with data augmentation techniques. This is meant to incorporate the so-called \textit{smoothness assumption} in SSL, which states that a classifier should be robust to local perturbations in its input. This is the case of the five methods explored in our work: MT, DCT, MM, RMM, and FM. \corrige{If we follow Van Engelen \textit{et al.}'s taxonomy, MT is a \textit{consistency regularization} method, in which predictions of a teacher and a student models are penalized when being different. DCT is described as a \textit{pseudo-labeling} method, based on the \textit{disagreement} between two models trained on two different views of the same data. As we shall see in the DCT description, the second view is automatically created by deriving adversarial examples of the original data samples. Finally, MM, RMM and FM are considered as \textit{hybrid} methods, in that they combine pseudo-labeling, consistency regularization and entropy minimization for performance improvement. Entropy minimization refers to methods that artificially lower the uncertainty of the predictions made on the unlabeled data. We will see, for instance, the use of a \textit{sharpening} function in MM.} 

\corrige{\subsection*{Semi-supervised deep learning in audio classification}}
In the seminal articles in which the five SSL methods were proposed, the experiments were carried out on image classification tasks only, not on audio related tasks. If we focus on SSL applied to sound event detection (SED), the most used technique in the literature is MT. In particular, the system ranked first in the Detection and Classification of Acoustic Scenes and Events (DCASE) task 4 2018 challenge (Large-scale weakly labeled semi-supervised  sound  event  detection  in  domestic  environments) used MT with convolutional recurrent neural networks trained on a small labeled subset and a larger unlabeled one~\cite{Lu2018}. Since then, MT was used in the baseline system provided by the challenge organizers, and most of the systems proposed by the participants~\cite{Turpault2019_DCASE,Miyazaki2020}. Also in the framework of DCASE Task 4, Shi and colleagues adapted MM for the task~\cite{Shi2019}. Their MM method outperformed their solution based on MT\footnote{http://dcase.community/challenge2019/task-sound-event-detection-in-domestic-environments-results}. SED is a task consisting of segmenting an audio recording in possibly overlapping audio events. It is slightly different from audio classification, the target task of the present work, in which we more simply aim to tag audio recordings globally with a single audio event category per recording. % Thus, to the best of our knowledge, the present work is the first one to compare five different SSL algorithms on an audio task, namely audio tagging.
Outside DCASE, MT has been favorably compared to supervised learning in~\cite{lu2019semi} for audio classification. The authors show the importance of using diverse collections of noise as perturbations in MT. They also used MixUp successfully, as we will in the present article. Although they used two datasets in common with us (Google Speech Commands and UrbanSound8k), their results cannot be compared to ours because of differences in the evaluation strategies: train/test splits different from the official ones and no cross-validation on UrbanSound8k, and a different number of target classes with Google Speech Commands. Finally, recently, FM and MT were compared on music, industrial sounds, and acoustic scenes classification data sets. FM outperformed MT and supervised learning in all cases~\cite{grollmisch2021improving}.

\corrige{\subsection*{An extension of our previous work}}
In previous work, we already compared two SSL methods for AT, namely MT and DCT, and we showed that DCT was consistently better than MT~\cite{cances2021comparison}. We build on this preliminary work to consider three simpler SSL methods, based on a single neural network instead of two models: MM, RMM and FM. Although some of these SSL methods were applied (in modified forms) to audio data in the context of audio classification before, as we just saw, the present work is among the first ones to compare a number of them in a systematic way. 

As we shall see in their technical description, a key aspect in these three ``hollistic'' methods is the extensive use of data augmentation techniques both on the labeled and unlabeled data subsets. In the results that we will report, we used the same augmentation techniques to train our fully-supervised baselines, which gave much stronger baselines than in our previous work~\cite{cances2021comparison}. Finally, another novelty of the present work is the addition of the mixup~\cite{zhang2018mixup} augmentation to the SSL methods MT, DCT and FM. % As we shall see, mixup led to consistent improvements.

% We can find audio applications of MT~\cite{tarvainen2018mean} in the Detection and Classification of Acoustic Scenes and Events (DCASE) task 4 challenges, namely the weakly supervised Sound Event Detection task. %The 2018 winners trained convolutional neural networks (CNN) on both a small labeled subset and a larger unlabeled one~\cite{Lu2018}

\section{Audio data augmentation}
\label{sec:aug_mixup}

Augmentations are at the heart of most recent semi-supervised learning mechanisms. In this section, we begin by describing the mixup mechanism, which we extensively use in this work, and the other audio data augmentations used in some of the training settings.

\subsection{Mixup}
\label{sec:mixup}

Mixup~\cite{zhang2018mixup} is a successful data augmentation/regula\-ri\-za\-tion technique, that proposes to mix pairs of samples (images, audio clips, etc.). If $x_1$ and $x_2$ are two different input samples (spectrograms in our case) and $y_1$, $y_2$ their respective one-hot encoded labels, then the mixed sample and target are obtained by a simple convex combination: 

\begin{align}
x^{mix} & = \lambda x_1 + (1 - \lambda)x_2 \\ \nonumber
y^{mix} & = \lambda y_1 + (1 - \lambda)y_2
\end{align}
where $\lambda$ is a scalar sampled from a symmetric Beta distribution at each mini-batch generation: 
% alpha -> 0 => lambda -> 0 or lambda -> 1, low perturbation, almost no mix
% alpha -> 1 => lambda ~ uniform, stronger mix

\begin{equation}
    \lambda \sim \text{Beta}(\alpha, \alpha)
\end{equation}
% \[\lambda \sim \text{Beta}(\alpha, \alpha)\]
where $\alpha$ is a real-valued hyper-parameter to tune (always smaller than 1.0 in our case).

In the original MM algorithm, an ``asymmetric'' version of mixup is used, in which the maximum value between $\lambda$ and $1-\lambda$ is retrieved:
\begin{equation}
    \lambda = \max(\lambda, 1 - \lambda)
\end{equation}
\corrige{This makes the $\lambda$ values either close to one, allowing the resulting mixed batches to be closer to $x_1$. This may be useful when the method mixes labeled and unlabeled samples, when only slight perturbations are wanted.}

\subsection{Audio signal augmentation methods}

\begin{table}[b]
    \caption{Augmentation hyperparameters.}
    \centering
    \label{table:augments}
    
    \begin{tabular}{@{}llcc@{}}
        \toprule
                   & Param. & Weak range & Strong range \\ \midrule
        Occlusion      & max size           & {[}0.25, 0.25{]}    & {[}0.75, 0.75{]}      \\
        CutOut     & scale               & {[}0.10, 0.50{]}      & {[}0.50, 1.00{]}    \\
        Speed perturb. & rate                & {[}0.50, 1.50{]}      & {[}0.25, 1.75{]}    \\
        % \midrule
        % Flip (RMM only)   & angle              & {[}0, 90, 180, 270{]} & {[}0, 90, 180, 270{]} \\ 
        \bottomrule
    \end{tabular}
\end{table}

We tested several audio augmentation techniques and retained three of them: \textit{Occlusion}, \textit{CutOut}~\cite{devries2017improved}, and \textit{Speed Perturbation}~\cite{ko2015audio}. 
In addition to the three selected augmentations described below, we also tried to add uniform noise on the log-mel spectrograms, invert the mel frequency axis and the time axis, but no gains were observed with these techniques. 

\begin{itemize}
    \item \textit{Occlusion}: applied to the raw audio signal, \textit{Occlusion} consists of setting a segment of the waveform to zero. The size of the segment is randomly chosen up to a user-defined maximum size. The position of the segment is also chosen randomly.
    
    % \item \textbf{Time Stretching (TS)~\cite{Salamon_2017}}: Also applied to the raw audio signal, Time Stretching consists of up-sampling or down-sampling the signal. The rate with which the signal is modified is chosen randomly within a predefined interval. The resulting augmented file is either shorter or longer. A padding or cropping is randomly applied at the start and the end of the stretched signal to keep the shape of the samples constant.
    
    \item \textit{CutOut}: applied to the log-mel spectrograms, CutOut sets the values within a random rectangle area with the -80 dB value, which corresponds to the silence energy level in our spectrograms. The length and width of the removed sections are randomly chosen from a predefined interval and depend on the spectrogram size.
    
    \item \textit{Speed Perturbation}: we resample the raw audio signal up (nearest-neighbor upsampling) or down (decimation) according to a rate chosen randomly within a predefined interval. The resulting waveform is either shorter or longer. Padding or cropping is randomly applied at the start and the end of the stretched signal to keep the signal duration constant. 

%    \item \textit{Flip} (auxiliary task of RMM only): Applied to the log-mel spectrograms, the orientation of the flip is randomly chosen and apply on the entire sample. \thomas{This augmentation is used in the .}
\end{itemize}

% \begin{figure}[h!]
%   \centering
%   \includegraphics[width=0.60\linewidth]{0.png}
%   \caption{Example of different augmentations on an audio file from UrbanSound8k, from top to bottom: original, weak and strong speed perturbation, and mixup augmentation.}
%   \label{fig:ex_aug}
% \end{figure}

The difference between Occlusion and CutOut is that CutOut sets a time-frequency rectangle to the -80 dB value, whereas Occlusion sets to zero a whole portion of the waveform. 

We used Occlusion, CutOut and Speed Perturbation in augmented supervised learning settings, and in MM, RMM, and FM. During training, one of those is randomly applied to each audio sample. % In the case of RMM, a fourth augmentation, here named \textit{Flip} is used in a regularization term, as we shall see here-after in Section~\ref{sec:rmm}. It makes use of random horizontal or vertical flips of the spectrograms. 

RMM and FM make use of so-called ``weak'' and ``strong'' augmentations. The difference between the two lies in the strength and randomness with which an augmentation is applied. \corrige{A ``weak'' augmentation has a 50\% chance to be applied, and a``strong'' one is always applied.}

In order to tune these augmentations, we performed a grid-search on their hyperparameters, training Wide-Resnet28-2 models on the Google Speech Commands dataset (this architecture and dataset will be described here-after). The resulting hyperparameters are listed in Table~\ref{table:augments}.

No augmentation was used in DCT nor in MT, except Gaussian noise in MT. % We tried to use the weak augmentation versions and no improvement was observed.

\section{Semi-supervised deep learning algorithms}
\label{sec:ssl}

This section provides a detailed description of the five SSL approaches we compare for audio classification. We chose them for their high performance reported for object recognition in images. Two of these approaches, Mean Teacher (MT)~\cite{tarvainen2018mean}, and Deep Co-Training (DCT)~\cite{qiao2018deep} use the principle of consistency regularization between the outputs of two models. The other methods, MixMatch~(MM)~\cite{berthelot2019mixmatch}, ReMixMatch~(RMM)~\cite{berthelot2020remixmatch}, and FixMatch~(FM)~\cite{sohn2020fixmatch}, use a single model and combine the three SSL mechanisms described in the introduction. 

We provide a figure to illustrate each of the five methods. In Section~\ref{subsec:addingmixup}, we explain how we add mixup to MT, DCT and FM, since MM and RMM already use it. We included  a blue box in the method workflow figures, to show where mixup is optionally integrated. We will refer to the modified methods as ``method+mixup'', for instance, FM+mixup.

\subsection{Mean Teacher (MT)}
\label{sec:mean_teacher}

\begin{figure}[ht!]
    % EL - EURASIP : fix LCC INPUT
    % \includegraphics[width=0.60\linewidth]{img/mean_teacher.png}
    \includegraphics[width=0.60\linewidth]{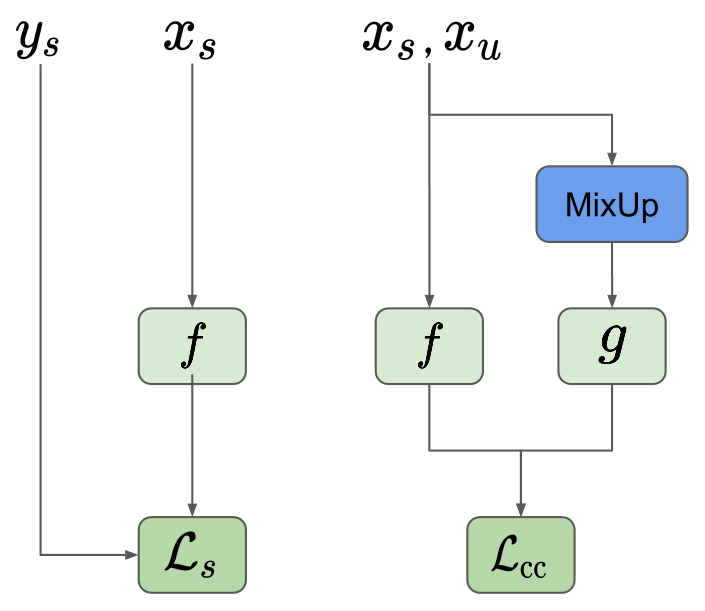}
    \centering
    % \caption{MT workflow. Both models receive as input labeled $x_s$ and unlabeled files $x_u$. A supervised loss $\mathcal{L}_s$ is computed between the ground truth and the student model predictions, whereas a consistency cost $\mathcal{L}_\mathrm{cc}$ is computed between the student and teacher model predictions. Mixup is added in later experiments. When not used, $y_s^{mix} = y_s$ and $x_s^{mix} = x_s$. The same goes for $x_u$.}
    \caption{MT workflow. Both models receive as input labeled $x_s$ and unlabeled files $x_u$. A supervised loss $\mathcal{L}_s$ is computed between the ground truth and the student model predictions, whereas a consistency cost $\mathcal{L}_\mathrm{cc}$ is computed between the student and teacher model predictions.}
    \label{fig:schema_mt}
\end{figure}

MT uses two neural networks: a ``student'' $f$ and a ``teacher'' $g$, which share the same architecture. The weights $\omega$ of the student model are updated using the standard gradient descent algorithm, whereas the weights $W$ of the teacher model are the Exponential Moving Average (EMA) of the student weights. The teacher weights are computed at every mini-batch iteration $t$, as the convex combination of its weights at $t\shortminus 1$ and the student weights, with a smoothing constant $\alpha_\mathrm{ema}$: % as shown in equation~\ref{eq:Wt}. The  controls the averaging time period.
%T: step t? epoch?

\begin{equation}
\label{eq:Wt}
    W_t=\alpha_\mathrm{ema} \cdot W_{t\shortminus 1} + (1 - \alpha_\mathrm{ema}) \cdot \omega_t
\end{equation}

There are two loss functions applied either on the labeled or unlabeled data subsets. On the labeled data $x_s$, the usual cross-entropy (CE) is used between the student model's predictions and the ground-truth $y_s$.

% \begin{equation}
%     \mathcal{L}_{\text{sup}}=\text{CE}(f(x_s), y_s)
% \end{equation}
% EL - EURASIP : fix lsup
\begin{equation}
    \mathcal{L}_{s}=\text{CE}(f(x_s), y_s)
\end{equation}

% Second, the consistency cost will be calculated from the predictions of the two models on unlabeled data. In our case, it is a Mean Square Error (MSE), and $x_u$ represent unlabeled data, $f(x_u)$ and $g(x_u)$ the prediction of the student and teacher model.

% The consistency cost is computed from the student prediction $f(x_u)$ and the teacher prediction $g(x_u')$, where $x_u$ is a sample from the unlabeled subset, and $x'_u$ the same sample but slightly perturbed with Gaussian noise and a 15 dB signal-to-noise ratio~\cite{Turpault2019_DCASE}. In our case, this cost is a Mean Square Error (MSE) loss:  % (replaced by mixup in the MT+mixup configuration).
% EL - EURASIP - Fix lcc input
\corrige{The consistency cost is computed from the student predictions $f(x_s)$ and $f(x_u)$, and from the teacher prediction $g(x_s')$ and $g(x_u')$, where $x_s'$ and $x_u'$ correspond to the same samples but slightly perturbed with Gaussian noise with a 15 dB signal-to-noise ratio~\cite{Turpault2019_DCASE}.} This cost is a Mean Square Error (MSE) loss:
% \begin{equation}
%     \mathcal{L}_{\text{cc}}=\text{MSE}(f(x_u), \perp g(x'_u))
% \end{equation}
% \corrige{
% \begin{equation}
%     \mathcal{L}_{\text{cc}}=\text{MSE}(f(x_s), \perp g(x_s'))+\text{MSE}(f(x_u), \perp g(x_u'))
% \end{equation}
% }
\corrige{
\begin{eqnarray}
    \mathcal{L}_{\text{cc}} & = & \text{MSE} ( f(x_s), \perp g(x_s') ) \nonumber\\
    && + \text{ } \text{MSE} (f(x_u), \perp g(x_u') )
    \label{eq:Lcc}
\end{eqnarray}
}
The symbol $\perp$ denotes the stop gradient operator, meaning that the teacher weights $W_t$ are a constant with respect to optimization.

The final loss function is the sum of the supervised loss function and the consistency cost weighted by a factor $\lambda_{cc}$ which controls its influence.

% \begin{equation}
% \label{eq:mt_loss}
%     \mathcal{L}_{\text{total}} = \mathcal{L}_{\text{sup}} + \lambda_{\text{cc}} \cdot \mathcal{L}_{\text{cc}}
% \end{equation}
% EL - EURASIP : fix lsup
\begin{equation}
\label{eq:mt_loss}
    \mathcal{L}_{\text{total}} = \mathcal{L}_{s} + \lambda_{\text{cc}} \cdot \mathcal{L}_{\text{cc}}
\end{equation}

\subsection{Deep Co-Training (DCT)}

\begin{figure}[h!]
  \centering
    \includegraphics[width=0.95\linewidth]{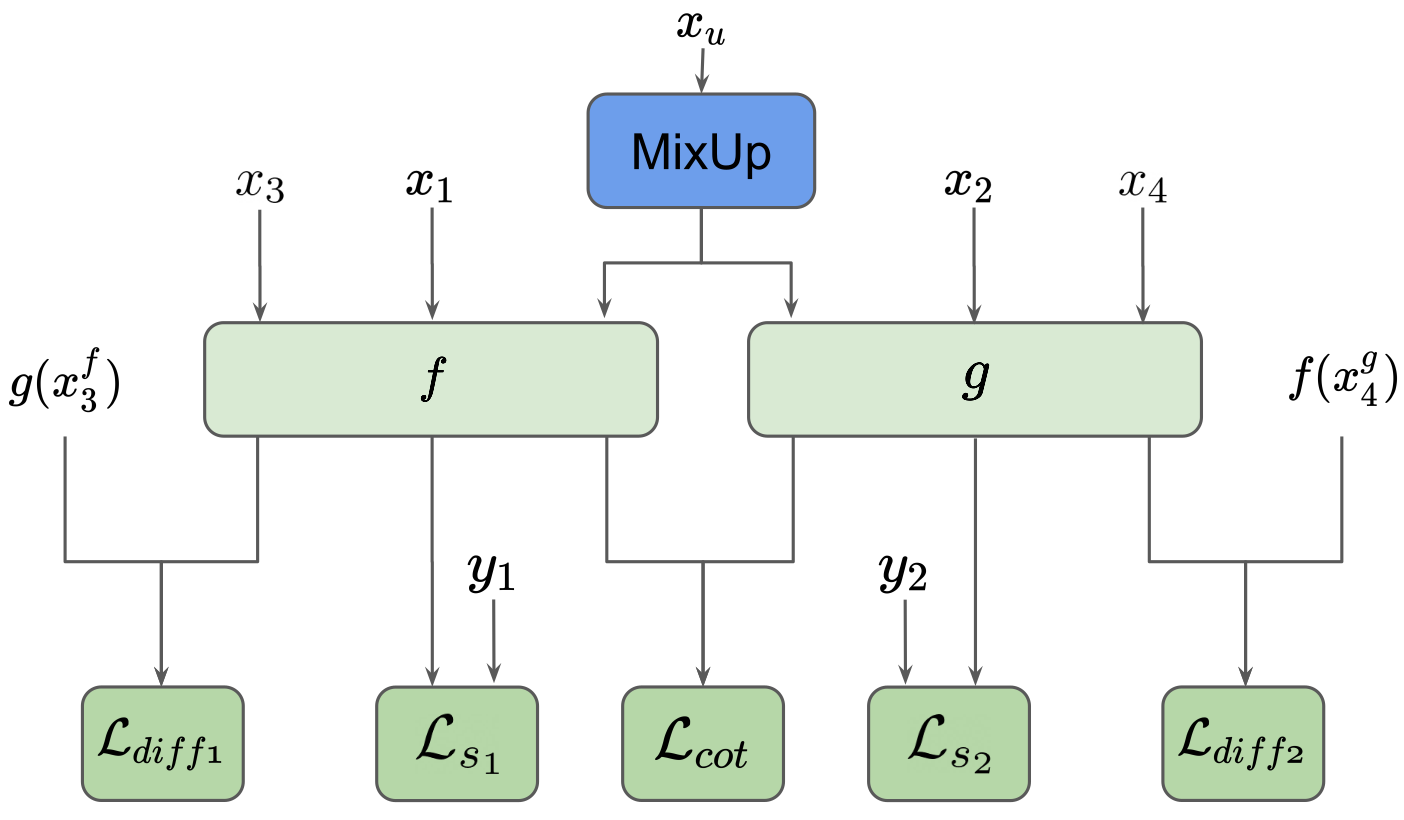}
    \caption{DCT workflow. Each model is trained on its own labeled samples $x_i$, unlabeled samples $x_u$ and the adversarial examples generated by the other model. Model $f$ makes predictions on $x_1$ and $x^g_4$, and model $g$ on $x_2$ and $x^f_3$. In our DCT+mixup variant,  mixup is used on the unlabeled samples only.}
  \label{fig:dct}
\end{figure}

DCT has been recently proposed by Qiao \textit{et al.}~\cite{qiao2018deep}. It is based on Co-Training (CT), the well-known generic framework for SSL proposed by Blum and colleagues in 1998~\cite{blum1998combining}. The main idea of Co-Training is based on the assumption that two independent views on a training dataset are available to train two models separately. Ideally, the two views are conditionally independent given the class. The two models are then used to make predictions on the unlabeled data subset. The most confident predictions are selected and added to the labeled subset. This process is iterative, like pseudo-labeling.

DCT is an adaptation of CT in the context of deep learning. Instead of relying on views of the data that are different, DCT makes use of adversarial examples to ensure the independence in the ``view'' presented to the models.
%The second difference is that the whole unlabeled dataset is used during training.
Each batch is composed of a supervised and an unsupervised part. Thus, the unlabeled data are directly used, and the iterative aspect of the algorithm is removed.

Let $\mathcal{S}$ and $\mathcal{U}$ be the subsets of labeled and unlabeled data, respectively, and let $f$ and $g$ be the two neural networks that are expected to collaborate. 

The DCT loss function is comprised of three terms, as shown in Eq.~(\ref{eq:L}). These terms correspond to loss functions estimated either on $\mathcal{S}$, $\mathcal{U}$, or both. Note that during training, a mini-batch is comprised of labeled and unlabeled samples in a fixed proportion. Furthermore, in a given mini-batch, the labeled examples given to each of the two models are sampled independently.  

% \begin{equation}
%     \mathcal{L} = \mathcal{L}_{\mathrm{sup}} + \lambda_{\mathrm{cot}}\mathcal{L}_{\mathrm{cot}} + \lambda_{\mathrm{diff}}\mathcal{L}_{\mathrm{diff}}
%     \label{eq:L}
% \end{equation}

\begin{equation}
    \mathcal{L} = \mathcal{L}_{s} + \lambda_{\mathrm{cot}}\mathcal{L}_{\mathrm{cot}} + \lambda_{\mathrm{diff}}\mathcal{L}_{\mathrm{diff}}
    \label{eq:L}
\end{equation}

% The first term, $\mathcal{L}_{\mathrm{sup}}$, given in Eq.~(\ref{eq:Lsup}), corresponds to the standard supervised classification loss function for the two models $f$ and $g$, estimated on examples $x_1$ and $x_2$ sampled from $\mathcal{S}$.
\corrige{
The first term, $\mathcal{L}_{s}$, given in Eq.~(\ref{eq:Lsup}), corresponds to the standard supervised classification loss function for the two models $f$ and $g$, estimated on examples $x_1$ and $x_2$ respectively, which are sampled from $\mathcal{S}$.
}

In our case, we use categorical Cross-Entropy (CE), the standard loss function used in classification tasks with mutually exclusive classes. 

% \begin{equation}
%     \mathcal{L}_{\mathrm{sup}} = \mathrm{CE}(f(x_1), y_1) + \mathrm{CE}(g(x_2), y_2)
%     \label{eq:Lsup}
% \end{equation}

\begin{equation}
    \mathcal{L}_{s} = \mathrm{CE}(f(x_1), y_1) + \mathrm{CE}(g(x_2), y_2)
    \label{eq:Lsup}
\end{equation}

As in MT, a consistency cost on the unlabeled examples is used in DCT. It takes the form of the Jensen-Shannon (JS) divergence between the two sets of predictions on examples $x_u$ sampled from the unlabeled subset $\mathcal{U}$, given by:
% Indeed, there is no need to minimize this divergence also on $\mathcal{S}$ since $\mathcal{L}_{\mathrm{sup}}$ already encourages the two models to have similar predictions on $\mathcal{S}$. 
% In SSL and Co-Training, the two classifiers are expected to provide consistent and similar predictions on both the labeled and unlabeled data. To encourage this behavior, the Jensen-Shannon (JS) divergence between the two sets of predictions is minimized on examples $x_u$ sampled from the unlabeled subset $\mathcal{U}$ only. Indeed, there is no need to minimize this divergence also on $\mathcal{S}$ since $\mathcal{L}_{\mathrm{sup}}$ already encourages the two models to have similar predictions on $\mathcal{S}$. Eq.~(\ref{eq:Lcot}) gives the JS analytical expression, with $H$ denoting entropy. 

\begin{eqnarray}
    \mathcal{L}_{\mathrm{cot}} & = & H\Big(\frac{1}{2}(f(x_u) + g(x_u))\Big) \nonumber\\
    &&{-}\: \frac{1}{2}\Big(H(f(x_u)) + H(g(x_u))\Big)
    \label{eq:Lcot}
\end{eqnarray}
where $H$ denotes the entropy. 

For DCT to work, the two models need to be complementary: on a subset different from $S\cup U$, examples misclassified by one model should be correctly classified by the other model~\cite{krogel2004multi}. In DCT, this is achieved by generating adversarial examples with one model and training the other model to be robust to these adversarial samples. To generate adversarial examples, we used the Fast Gradient Signed Method (FGSM, \cite{goodfellow2015explaining}), as in Qiao's work. The $\mathcal{L}_{\mathrm{diff}}$ loss term (Eq.~(\ref{eq:Ldiff})) is the sum of the Cross-Entropy losses between the predictions $f(x_3)$ and $g(x^{f}_3)$, where $x_3$ is sampled from $S\cup U$ and $x^{f}_3$ is the adversarial example generated with the model $f$ from $x_3$ taken as input. \corrige{The second term is the symmetric term for model $g$, with $x_4$ sampled from $S\cup U$ and $x^{g}_4$ the adversarial example generated with $g$ from $x_4$.}
\begin{eqnarray}
    \mathcal{L}_{\mathrm{diff}} & = & \mathrm{CE}(f(x_3), g(x^{f}_3)) \nonumber \\
    && {+}\: \mathrm{CE}(g(x_4), f(x^{g}_4))
    \label{eq:Ldiff}
\end{eqnarray}

For more in-depth details on the technical aspects of DCT, the reader may refer to~\cite{qiao2018deep}. We implemented DCT as precisely as described in Qiao's article, using PyTorch, and made sure to accurately reproduce their results on CIFAR-10: about 90\% accuracy when using only 10\% of the training data as labeled data (5000 images).

\subsection{MixMatch}

\begin{figure}[h!] 
    \includegraphics[width=0.65\linewidth]{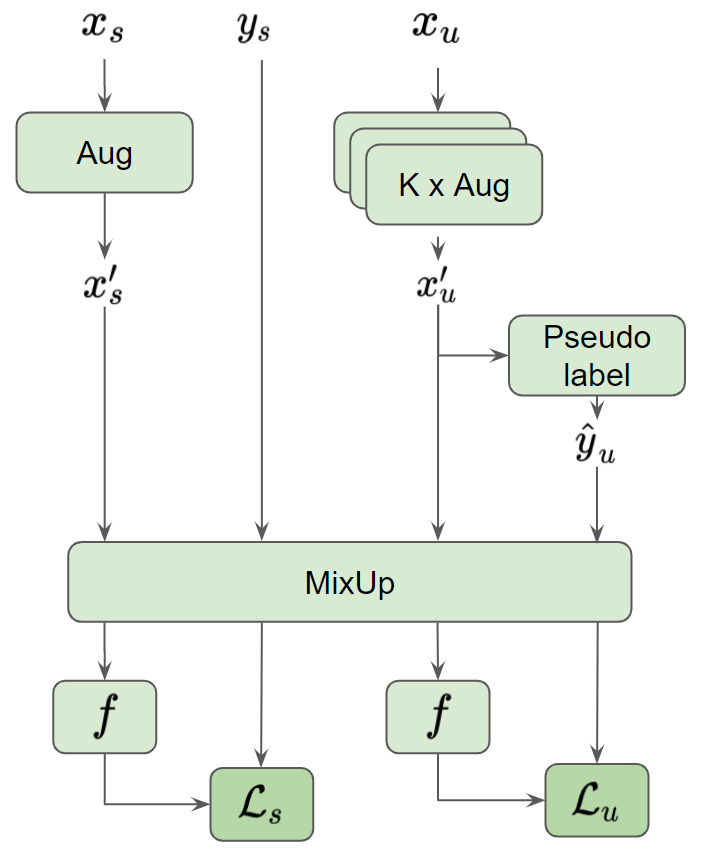}
    \centering
    \caption{MM workflow. $K$ augmentations are applied to the unlabeled data $x_u$, and the averaged model predictions are used as pseudo labels $\hat{y}_u$. The labeled and augmented unlabeled data are mixed up and used to compute the supervised and unsupervised loss values.}
    \label{fig:schema_mixmatch}
\end{figure}

% lc - The MixMatch method \cite{berthelot2019mixmatch} is an holistic technique that use the benefits of mixup \cite{zhang2018mixup}, Pseudo-Labeling and data augmentations for semi-supervised training. The method needs only one network which will be used for guessing a pseudo-label on augmented unlabeled data.  Then the labeled and unlabeled data is mixed with a mixup algorithm for improve the generalization of the network on training data.

MixMatch~\cite{berthelot2019mixmatch} (MM) uses entropy minimization and standard regularization, namely pseudo-labeling~\cite{lee2013}, mixup, and weak data augmentation, to leverage the unlabeled data and provide better generalization capabilities. Unlike MT and DCT, this approach uses only one model. The different steps are shown in Fig.~\ref{fig:schema_mixmatch} and detailed in the following paragraphs.

During the learning phase, each minibatch is composed of labeled $x_s$ and unlabeled  $x_u$ samples in equivalent proportions. The first step consists of applying an augmentation to the labeled part of the mini-batch and $K$ augmentations to the unlabeled part in parallel. \corrige{These $K$ augmentations are sampled from the three augmentations (weak) described in Section~\ref{sec:aug_mixup}}. In the second step, pseudo-labels $y_u$ are generated for the unlabeled files using the model's prediction averaged on these $K$ variants as shown in Eq.~(\ref{e:xu_weak}), where $x'_{u,i}$ denotes the $i$-th variant of an unlabeled augmented file.

% lc
% The MixMatch algorithm starts by getting a labeled batch $X_s$, the corresponding labels $Y_s$ and an unlabeled batch $X_u$. Then the weak augmentation is applied to labeled batch one time and to the unlabeled batch k times for having k different variations $X'_{u,i}$ of the same unlabeled batch.
% \begin{align}
%     \forall x_s \in X_s, x'_s & = \text{weak}(x_s) \\
%     \forall i \in 1..k, \forall x_u \in X_u, x'_{u, i} & = \text{weak}(x_u)
% \end{align}

% Then we use the k augments of $X_u$ to guess the labels on unlabeled data by compute the average of predictions on the k variants :
\begin{equation}
    \label{e:xu_weak}
    % \forall x'_u \in x'_u, \hat{y}_u = \frac{1}{k} \sum_{i=1}^k f(x'_{u, i})
    \hat{y}_u = \frac{1}{k} \sum_{i=1}^k f(x'_{u, i})
\end{equation}

% lc - This pseudo-label can have a high entropy, which means it can be close to an uniform distribution. For encouraging the model to produce confident output, we decrease the entropy of this label by applying a post-processing function "sharpen", defined in equation~\ref{e:sharpen}:

For encouraging the model to produce confident predictions, a post-processing step is necessary to decrease the output's entropy. To do so, the highest probability is increased and the other ones decreased. This process is called "sharpening" by the method authors, and it is defined as:

\begin{equation}
    \label{e:sharpen}
    % \forall i \in 1..|\hat{y}|, \text{sharpen}(\hat{y}_i, T) = \frac{\hat{y}_i^{1 / T}}{\sum_{j=1}^{|\hat{y}|}
    \text{sharpen}(p, T)_i := p_i^{1 / T} \Bigg/ \sum_{j=1}^{|p|} p_j^{1 / T}
\end{equation}

The sharpen function is applied on to the pseudo-labels $p = \hat{y}_u$. The parameter $T$, called Temperature, controls the strength of the sharpen function. When T tends towards zero, the entropy of the distribution produced is lowered. 
% lc - This function increases the higher probability of $y$ and reduce the other values. 

% lc - Then we concatenate the labeled and unlabeled batch on the first axis and mix it with labeled and unlabeled data :
% \begin{align}
%     W & = \text{Concat}({X'}_s, {X'}_u) \\
%     Q & = \text{Concat}(Y_s, \hat{Y}_u) \\
%     W, Q & = \text{Shuffle\_Order}(W, Q) \\
%     {X'}_s^{\text{mix}}, Y_s^{\text{mix}} & = \text{mixup}({X'}_s, W_{1..B_s}, Y_s, Q_{1..B_s}) \\
%     {X'}_u^{\text{mix}}, \hat{Y}_u^{\text{mix}} & = \text{mixup}({X'}_u, W_{(B_s+1)..(B_s+B_u)}, \\
%     & \hat{Y}_u, Q_{(B_s+1)..(B_s+B_u)})
% \end{align}

Finally, the labeled and unlabeled augmented samples are concatenated and shuffled into a $W$ set then used as a pool of training samples used by the asymmetric mixup function. Asymmetric mixup is applied separately on the labeled and unlabeled parts of the mini-batch, as formulated here:
\begin{equation}
    \label{e:x_mix_s}
    {x'}^\mathrm{mix}_s = \text{mixup}(x_s | W_{1\ldots B_s})
\end{equation}
\begin{equation}
    \label{e:x_mix_u}
    {x'}^\mathrm{mix}_u = \text{mixup}(x_u | W_{|x_s|+1\ldots|W|})
\end{equation}

% lc - The $Shuffle\_Order$ operation shuffle in a same random order the batch W and the labels Q. The batch ${X'}_s^{mix}$ contains labeled data perturbed by other labeled or unlabeled data. This mix is closer to labeled data due to the asymmetric mixup operation (see~\ref{sec:mixup}). In the same way, the second batch ${X'}_u^{mix}$ contains unlabeled data perturbed by labeled or unlabeled data that was not used to perturb the other mixed batch. \\

where $B_s$ and $|W|$ are the number of labeled samples and of the whole W set. The W set and the corresponding labels are shuffled in the same order. Each labeled sample is then perturbed by a second labeled or unlabeled sample. Mixing the two is done so that the original labeled sample remains the main component of the resulting sample. The operation has been detailed in Section~\ref{sec:mixup}. The same procedure is applied onto the unlabeled files using the remaining samples from W.

% lc - The original MixMatch loss function is composed by a standard Cross-Entropy for the supervised loss component $\mathcal{L}_s$ and a squared $\ell _2$ norm for the unsupervised loss component $\mathcal{L}_u$. In all our experiments, we replace the squared $\ell _2$ norm by a cross-entropy like proposed in ReMixMatch and FixMatch papers.

The original MixMatch loss function is composed of the standard CE cost for the supervised loss $\mathcal{L}_s$, and the MSE for the unsupervised loss $\mathcal{L}_u$. We replace MSE with CE in all our experiments, as proposed in the ReMixMatch paper. Indeed, it seems that CE performs better than MSE in our experiments.

\begin{equation}
    \mathcal{L}_s = \frac{1}{B_s} \sum_{({x'}_s^{\text{mix}}, y_s^{\text{mix}})} \text{CE}(f({x'}_s^{\text{mix}}), y_s^{\text{mix}})
\end{equation}

\begin{equation}
    \mathcal{L}_u = \frac{1}{K \cdot B_u} \sum_{({x'}_u^{\text{mix}}, \hat{y}_u^{\text{mix}})} \text{CE}(f({x'}_u^{\text{mix}}), \hat{y}_u^{\text{mix}})
\end{equation}
% \mathcal{L}_u =  \frac{1}{k \cdot Bu} \sum_{(x'_u^{mix}, y_u^{mix}) \in X'_u * Y_u^{mix}} || f(x'_u^{mix}) - y_u^{mix} ||_2^2
where $B_s$ and $B_u$ are the number of examples in the labeled and unlabeled mini-batches.

The final loss is the sum of the two components, with a hyper-parameter $\lambda_u$ :
\begin{equation}
    \mathcal{L} = \mathcal{L}_s + \lambda_u \cdot \mathcal{L}_u
\end{equation}

\subsection{ReMixMatch (RMM)}
\label{sec:rmm}

\begin{figure}[h!] % b
    % EL - EURASIP - Update RMM schema & caption : remove self-loss, fixed xui not in mixup !!
    % \includegraphics[width=0.97\linewidth]{img/RMM_vertical-reduce.png}
    \includegraphics[width=0.97\linewidth]{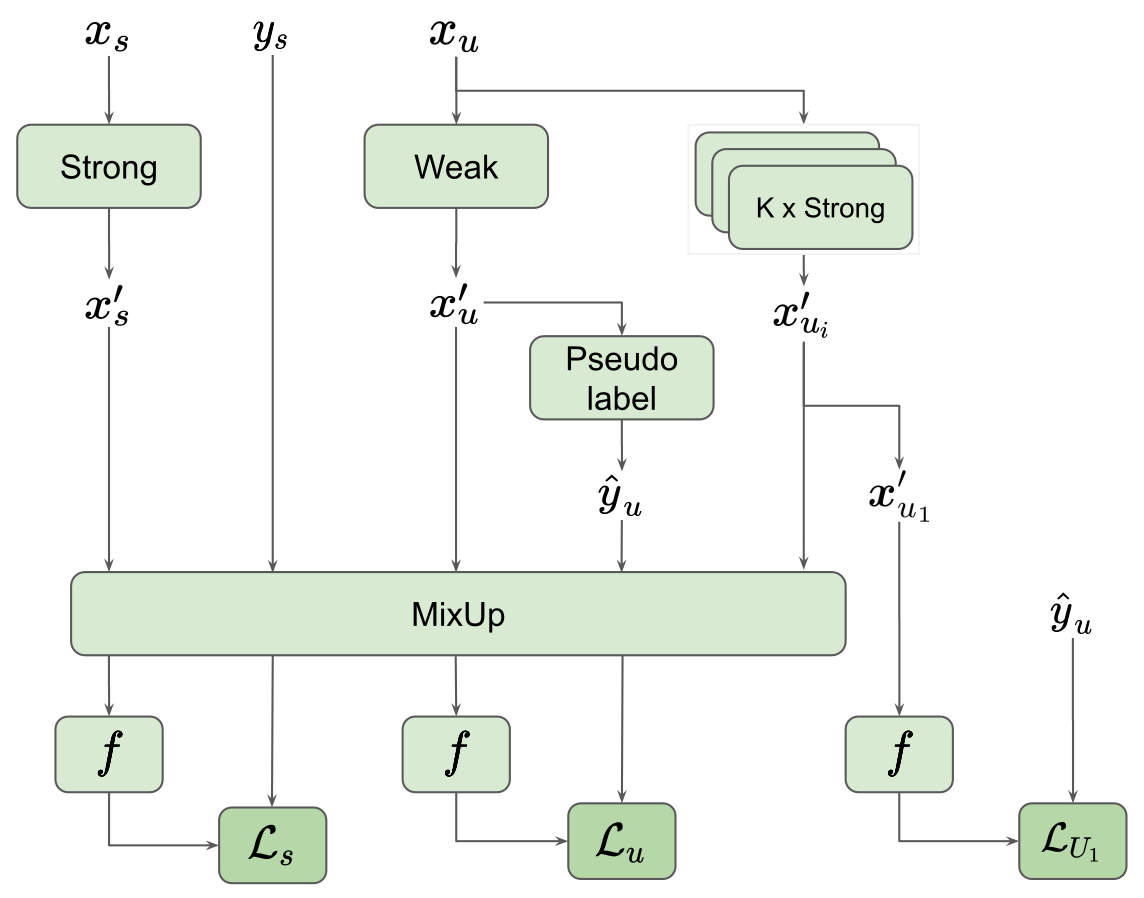}
    \centering
    \caption{\corrige{RMM workflow. One weak and $K$ strong augmentations are applied to the unlabeled data $x_u$. The weakly augmented unlabeled data $x'_u$ are used to create pseudo labels $\hat{y}_u$. The first batch of strongly augmented unlabeled data $x'_{u_i}, i=1$ is used in the unsupervised loss component $L_{u_1}$ (using the pseudo-labels $\hat{y}_u$).}}
    % \caption{RMM workflow. One weak and $K$ strong augmentations are applied to the unlabeled data $x_u$. The weakly augmented unlabeled data $x'_u$ are used to create pseudo labels $\hat{y}_u$. The first batch of strongly augmented unlabeled data $x'_{u_i}, i=1$ is used in the unsupervised loss component $L_{u_1}$ (using the pseudo-labels $\hat{y}_u$) and in the self-supervised loss $L_r$. In the self-learning component, the $y_r$ are binary labels, indicating either an horizontal or a vertical flip to be predicted.}
    \label{fig:schema_remixmatch}
\end{figure}

% ReMixMatch~(RMM)~\cite{berthelot2020remixmatch} is presented as an improvement of MixMatch and introduces the concept of strong and weak augmentations. It also adds a distribution alignment mechanism and a self-supervised loss component $\mathcal{L}_{r}$. 
% EL - EURASIP - remove self-loss in the desc ?
\corrige{ReMixMatch~(RMM)~\cite{berthelot2020remixmatch} was presented as an improvement of MixMatch and introduced the concept of strong and weak augmentations and a so-called distribution alignment mechanism.}

At every iteration, the batch is composed of labeled $x_s$ and unlabeled $x_u$ samples. One weak augmentation and $K$ strong augmentations are applied on $x_u$. The weakly-augmented sample is used to compute the pseudo-label vectors $\hat{y}_u$ of the unlabeled examples.

\begin{equation}
    \hat{y}_u = f\big(\text{weak}(x_u)\big)
\end{equation}

% EL - EURASIP reviewer #1, 13)
% The distribution alignment modifies the pseudo-label to follow the class distribution of the labeled part. In order to apply this process, we compute the labeled distribution $p_s$ with true labels and the unlabeled distribution $p_u$ with the previous pseudo-labels. Then we apply the distribution alignment on $\hat{y}_u$:
% notation N ??, keep equations 17 & 18 ?
\corrige{A distribution alignment mechanism modifies the pseudo-labels to make them follow the class distribution of the labeled subset. Two ``distributions'' $p_s$ and $p_u$ are estimated in the form of vectors, which are respectively the averages of the true labels $y_s$ and of the pseudo-labels $\hat{y}_u$, calculated over the samples of the $N$ previous batches. Then, distribution alignment is applied to $\hat{y}_u$ with this equation:}
% \begin{equation}
%     \hat{p}_{s} = \frac{1}{N} \sum_{i=t-N}^{t} y_{s,i}
% \end{equation}
% \begin{equation}
%     \hat{p}_{u} = \frac{1}{N} \sum_{i=t-N}^{t} \hat{y}_{u,i}
% \end{equation}
\begin{equation}
    \hat{y}_u = \text{Normalize}(\hat{y}_u \cdot \frac{p_s}{p_u})
\end{equation}
% In particular, we compute the mean of the last 128 batches labels to compute $p_s$ and the last 128 batches pseudo-labels to compute $p_u$.
Finally, we apply the sharpen function from Eq.~(\ref{e:sharpen}) to the pseudo-labels $\hat{y}_u$, as done in MixMatch.
The labels $\hat{y}_u$ will be used as targets for the weakly and strongly augmented batches. Like in MixMatch, we concatenate the labeled and unlabeled batches to a set $W$ for the mixup augmentation, and the labeled and unlabeled loss $\mathcal{L}_s$ and $\mathcal{L}_u$ remain the same.

ReMixMatch also introduced a strong-augmentation loss component for increasing stability and accuracy. This component will be computed with the first strongly-augmented version of $x_u$, called ${x'}_{u_1}$:
\begin{equation}
    \label{e:rmm_Lu1}
    \mathcal{L}_{u_1} = \frac{1}{B_u} \sum_{({x'}_{u_1}, \hat{y}_u)} \text{CE}(f({x'}_{u_1}), \hat{y}_u)
\end{equation}

\corrige{In the original ReMixMatch, the authors added another loss term, a self-supervised learning component that predicts which transformation is applied to the ${x'}_{u_1}$ batch. The transformation used was a rotation of 0, 90, 180, or 270 degrees, and the model had to guess which angle the image had been rotated by (a four-class classification task). In some configurations, it was supposed to help the model to avoid collapsing during training.  
% EL - EURASIP - phrase pour indiquer que la loss a été supprimée
This component was removed because it did not show any positive impact on our experiments, and using rotations or flips on audio spectrograms is difficult to justify in terms of audio semantics.
% Here, we choose to replace this component with a random flip strategy on the audio file spectrogram. The four transformations we used are: identity (i), vertical flip (v), horizontal flip (h) and horizontal+vertical flip (hv). The model tries to predict which flip is applied with an additional linear layer at the end of the model. This output is called $f_r$. Like the other components, the self-supervised component uses the standard cross-entropy:
% \begin{equation}
%     \label{e:rmm_Lr}
%     \mathcal{L}_{r} = \frac{1}{B_u} \sum_{({x'}_{u_1}, \hat{y}_u)} \text{CE} \Big( f_r \big( \text{flips}({x'}_{u_1}, y_r) \big), y_r \Big)
% \end{equation}
% where $y_r$ is sampled from $\{$i, v, h, hv$\}$.

In our experiments, the final loss is the sum of the three different components:
% \begin{equation}
%     \mathcal{L} = \mathcal{L}_s + \lambda_u \cdot \mathcal{L}_u + \lambda_{u_1} \cdot \mathcal{L}_{u_1} + \lambda_r \cdot \mathcal{L}_r
% \end{equation}
\begin{equation}
    \mathcal{L} = \mathcal{L}_s + \lambda_u \cdot \mathcal{L}_u + \lambda_{u_1} \cdot \mathcal{L}_{u_1}
\end{equation}
}

\subsection{FixMatch}

\begin{figure}[h!]
    \includegraphics[width=0.9\linewidth]{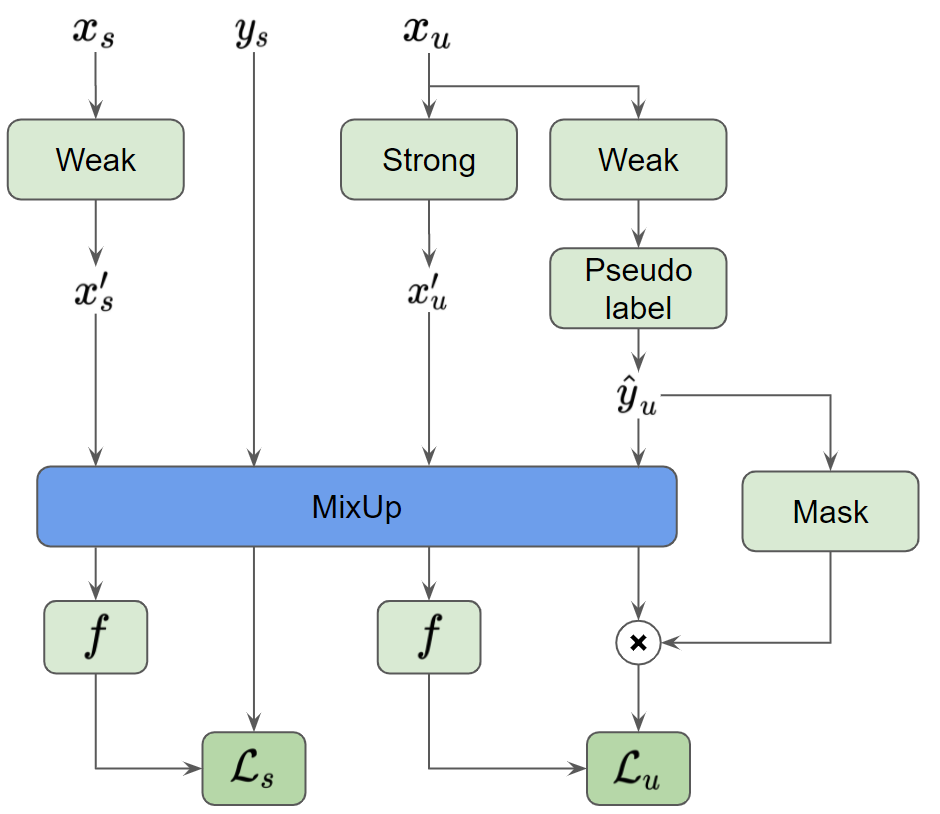}
    \centering
    \caption{FixMatch workflow. A weakly augmented version of $x_u$ is used to compute a pseudo-label $\hat{y}_u$ and a mask. The strongly augmented variant is used to compute the unlabeled loss term. The mixup component is used on a concatenated set of labeled and unlabeled samples (FixMatch+mixup).}
    \label{fig:schema_fixmatch}
\end{figure}

FixMatch~\cite{sohn2020fixmatch} (FM) is another SSL method which proposes a simplification of MM and ReMixMatch. The method also uses one model, removes mixup and replaces the sharpen function by binary pseudo-labels. FM uses both weak augmentations (weak) and strong augmentations (strong). The strong augmentations can mislead the model predictions by disrupting too much the training data. Figure~\ref{fig:schema_fixmatch} shows the main pipeline of FixMatch. As in the other method illustrations, we added a mixup box in blue, to indicate where we add it to the algorithm in our modified FM algorithm, thus called FM+mixup.

The supervised loss component is the standard cross-entropy applied to the weakly-augmented data :
\begin{equation}
    \mathcal{L}_s = \text{CE}\Big(f\big(\text{weak}(x_s)\big), y_s\Big)
\end{equation}

Then, we guess the labels of the weakly augmented unlabeled data and apply a binarization (argmax) of these predictions to have a one-hot encoded label. This label is used as target for training the model with strongly augmented unlabeled data. It allows the model to generalize with weak and strong augmentations and it also uses the guessed label to improve the model accuracy with unlabeled data:

\begin{equation}
    \hat{y}_u = f\big(\text{weak}(x_u)\big)
\end{equation}

% lc - In order to avoid training on incorrect guessed labels, FixMatch uses a threshold $\tau$. It removes the unsupervised loss if the higher value of the guessed label before binarization is below the threshold.
To avoid training on incorrect guessed labels, FM uses a threshold $\tau$ that ensures that the unsupervised cost function can only be applied to predictions made with high confidence, i.e., above this threshold. This can be easily implemented in the form of a mask:

\begin{align}
    \mathrm{mask}  & = \mathbbm{1} \big( \text{max}(\hat{y}_u) > \tau \big) \\\nonumber
    \mathcal{L}_u  & = \mathrm{mask} \cdot \text{CE} \Big( f\big(\text{strong}(x_u)\big), \text{argmax}(\hat{y}_u) \Big) 
\end{align}

As in MixMatch, we sum the loss components to compute the final loss:

\begin{equation}
    \mathcal{L} = \mathcal{L}_s + \lambda_u \cdot \mathcal{L}_u
\end{equation}

% {\color{blue}
% \subsection{UDA}
% Unsupervised Data Augmentation~\cite{xie2020unsupervised} (UDA) is another SSL method very similar to FM, originally tested on image and text classification tasks. It uses strong augmentations (RandAugment or Back-translation) on the unlabeled data, and a softmax-sharpening to post-process the pseudo-labels.

% The pseudo-labels are estimated on the original input data $x_u$, without augmentation:

% \begin{equation}
%     \hat{y}_u = \text{exp} \big( z / T \big) \Bigg/ \sum_i \text{exp} \big( z_i / T \big)
% \end{equation}

% Where $z$ is the logit output of the model $f$ for the unlabeled batch $x_u$ and $T$ a temperature hyperparameter which plays a role similar to the temperature of the MixMatch sharpen function. When the temperature tends to zero, the predictions have a lower entropy. 

% The supervised loss component $\mathcal{L}_s$ is a standard cross-entropy between the non-augmented data and its label and the unsupervised loss component is almost the same than in FixMatch, except that we do not binarize the pseudo-label:
% \begin{align*}
%     \mathrm{mask}  & = \mathbbm{1} \big( \text{max}(\hat{y}_u) > \tau \big) \\\nonumber
%     \mathcal{L}_u  & = \mathrm{mask} \cdot \text{CE} \Big( f\big(\text{strong}(x_u)\big), \hat{y}_u \Big)
% \end{align*}

% \noindent The final loss is the same as the MixMatch one in Eq~\ref{e:mm_final_loss}.
% }

\subsection{Adding mixup to MT, DCT and FM\label{subsec:addingmixup}}

As we described here-above, MM and RMM already uses mixup in its workflow. In order to measure the impact of mixup, we will  report results when we remove mixup from MM and RMM. On the contrary, the three other SSL methods explored in our work (MT, DCT, FM) do not use mixup in their original version. We explored several ways to add mixup to them, and retained the best one for each of the three methods. Note that we illustrate where the mixup operation has been added in the figures describing the different methods in the previous section.

Since the labeled and unlabeled data flow is very similar in MM and FM, we added mixup to FM at the same place as in MM: both labeled and unlabeled samples are mixed up. Similarly, it is also the asymmetric mixup variant that we used in MM and FM since mixup is applied to labeled and unlabeled samples together, as in the original MM method. Using mixup on labeled and unlabeled examples separately seems to hurt performance with these two methods. 

In MT, mixup is applied on labeled and unlabeled samples separately and only for the teacher model. The perturbation with Gaussian noise applied to the unlabeled samples is removed, since no gain was observed when mixup is used instead.

For DCT, mixup is applied on the unlabeled samples only, common to both models in each minibatch during training. Applying mixup on the labeled samples, which are sampled differently for the two models at each training step, lead yo worse results. It is then, not necessary to use the asymmetrical variant for MT and DCT.

Finally, in all cases, we apply mixup on the log-mel spectrograms, which are the input features given to our deep neural networks (feature extraction is detailed in the experiment section).

\section{Experiments}
\label{sec:exp}

In this section, we describe our experimental setup. We give a brief description of the datasets and metrics, describe the Wide ResNet architecture we used, together with the training strategy details.
% The experiments were carried out in two stages. First, each SSL method was applied on the three datasets and compared to the best supervised results obtained in the same configurations, i.e., the same training parameters and the same amounts of labeled files. % a supervised training using only weak augmentations (weak) for ESC-10, and weak augmentations with mixup (Weak+mixup) for UBS8K and GSC. The objective is to evaluate the efficiency of the SSL method versus the use of augmentation only.
% Second the significant impact of mixup performance in MixMatch~\cite{berthelot2019mixmatch} motivated us to evaluate its impact on the other supervised methods. Therefore, we decided to add this technique to DCT, MT, and FM.

\subsection{Datasets and evaluation metrics}
% We worked on three benchmark audio datasets. %Each of the datasets is different in size and content.

\label{sec:datasets}
\textbf{Environmental Sound Classification 10 (ESC-10)~\cite{esc-dataset}} is a selection of 400 five-second-long recordings of audio events separated into ten balanced categories. The dataset is provided with five uniformly sized cross-validation folds that will be used to perform the evaluation. The files are sampled at 44 kHz and are converted into $431\times64$ log-mel spectrograms.

\textbf{UrbanSound8k (UBS8K)~\cite{ubs8k-dataset}} is a dataset composed of 8742 files between 1 and 4 seconds long, separated into ten balanced categories. The dataset is provided with ten cross-validation folds of uniform size that will be used to perform the evaluation. The files are zero-padded to 4 seconds, resampled to 22 kHz, and converted to $173\times64$ log-mel spectrograms.

\textbf{Google Speech Commands Dataset v2 (GSC) \cite{speechcommandsv2}} is an audio dataset of spoken words designed to evaluate keyword spotting systems. The dataset is split into 85511 training files, 10102 validation files, and 4890 testing files. The latter is used for the evaluation of our systems. We ran the task of classifying the 35 word categories of this dataset. The files are zero-padded to 1 second if needed and sampled at 16 kHz before being converted into $32\times64$ log -mel spectrogram.

In all cases, the 64 mel-coefficients were extracted using a window size of 2048 samples and a hop length of 512 samples. For ESC-10 and UBS8K, we used the official cross-validation folds. We report the average classification Error Rate (ER) along with standard deviations. \corrige{ER is defined as the number of errors divided by the total number of samples.} %  We calculate ER as $1 \text{ - accuracy}$.

\subsection{Models}
\label{ssec:models}
% \subsubsection{Wideresnet28-2}
We used the Wide-ResNet-28-2~\cite{zagoruyko2017wide} architecture in all our experiments. This model is very efficient, achieving SOTA performance on the three datasets when trained in a $100\%$ supervised setting. Moreover, its small size, comprised of about 1.4 Million parameters, allows to experiment quickly.
Its structure consists of an initial convolutional layer (\texttt{conv1}) followed by three groups of residual blocks (\texttt{block1}, \texttt{block2}, and \texttt{block3}). Finally, an average pooling and a linear layer act as a classifier. The residual blocks, composed of two $\mathrm{BasicBlock}$, are repeated three times and their structure is defined in Eq.~(\ref{eq:Residual}). The number of channels of the convolution layers is referred as $l$, $\text{BN}$ stands for Batch Normalization and ReLU~\cite{agarap2019deep} for the Rectified Linear Unit activation function. We used the official implementation available in PyTorch~\cite{NIPS2019_9015}.

\begin{equation}
\label{eq:Residual}
\mathrm{BasicBlock}(l)=\left(
    \text{conv 3}\times\text{3}~@~l, 
    \text{BN}, 
    \text{ReLU} \right)
\end{equation}

\begin{table}[t]
    \begin{tabular}{@{}crcl@{}}
    \toprule
    Layer                   & \multicolumn{1}{c}{} & Architecture    & \multicolumn{1}{c}{}     \\ \midrule
    input                   & \multicolumn{3}{c}{Log mel spectrogram}                           \\ \midrule
    conv1                   & \multicolumn{1}{c}{} & BasicBlock(32)  & \multicolumn{1}{c}{}     \\ \midrule
    \multicolumn{1}{l}{}    & \multicolumn{1}{l}{} & Max pool        &                          \\ \midrule
    \multirow{2}{*}{block1} & \multirow{2}{*}{\Big{[}} & BasicBlock(32)  & \multirow{2}{*}{\Big{]} $\times$ 4} \\
                            &                      & BasicBlock(32)  &                          \\ \midrule
    \multirow{2}{*}{block2} & \multirow{2}{*}{\Big{[}} & BasicBlock(64)  & \multirow{2}{*}{\Big{]} $\times$ 4} \\
                            &                      & BasicBlock(64)  &                          \\ \midrule
    \multirow{2}{*}{block3} & \multirow{2}{*}{\Big{[}} & BasicBlock(128) & \multirow{2}{*}{\Big{]} $\times$ 4} \\
                            &                      & BasicBlock(128) &                          \\ \midrule
    \multicolumn{1}{l}{}    & \multicolumn{1}{l}{} & Avg pool        &                          \\ \midrule
    \multicolumn{1}{l}{}    & \multicolumn{1}{l}{} & ReLU            &                          \\ \midrule
                          & \multicolumn{1}{l}{} & Linear          &                          \\ \bottomrule
    \end{tabular}
    \centering
    \vspace{0.3cm}
    \caption{Architecture of Wide ResNet28-2. Downsampling is performed by the first layers in block2 and block3.}
\end{table}

\subsection{Training configurations}

Each model was trained using the ADAM~\cite{kingma2017adam} optimizer. Table~\ref{t:params} shows the  hyper-parameter values used for each method, such as the learning rate $\textrm{lr}$, the mini-batches' size $\textrm{bs}$, the warmup length $\textrm{wl}$ if used, and the number of epochs $e$. These parameters are identical regardless of the dataset used, unless otherwise specified. They were obtained by performing a reasonable short grid-search using UBS8K dataset first validation fold.

% EL - EURASIP - Fix learning it -> learning epoch & fix "e" in eq, (which means nb epochs) notation number of epochs : N_e
For supervised training, MM and FM, the learning rate remains constant throughout training. For MT and DCT, the learning rate is weighted by a descending cosine rule, function of the learning epoch $t$:

\begin{equation}
    \mathrm{lr} = 0.5 \Big( 1.0 + \text{cos}\big( (t \shortminus 1) \frac{\pi}{N_e}\big)\Big)
\end{equation}
\corrige{where $N_e$ denote the number of epochs.}

% TABLE HPARAMS
\begin{table}[ht]
    \centering
    \caption{Training parameters used on the datasets. Bs: batch size, lr: learning rate, wl: warm-up length in epochs, $N_e$: number of epochs, $\alpha$: mixup Beta param.}
    % \vspace{5mm}
    
    \rowcolors{2}{gray!25}{white}
    
    \begin{tabular}{l|ccccc}
        & bs & lr & wl & $N_e$ & $\alpha$ \\
        \midrule
        Supervised & 256 & 0.001 & - & 300 & - \\
        mixup      & 256 & 0.001 & - & 300 & 0.40 \\
        MT         & \hphantom{1}64 & 0.001 & \hphantom{1}50  & 200 & - \\
        MT+mixup   & \hphantom{1}64 & 0.001 & \hphantom{1}50  & 200 & 0.40 \\
        DCT        & \hphantom{1}64 & 0.0005 & 160 & 300 & - \\
        DCT+mixup  & \hphantom{1}64 & 0.0005 & 160 & 300 & 0.40 \\
        MM-mixup   & 256 & 0.001 & - & 300 & - \\
        MM         & 256 & 0.001 & - & 300 & 0.75 \\ 
        RMM-mixup  & 256 & 0.001 & - & 300 & - \\
        RMM        & 256 & 0.001 & - & 300 & 0.75 \\
        FM         & 256 & 0.001 & - & 300 & - \\
        FM+mixup   & 256 & 0.001 & - & 300 & 0.75 \\
    \end{tabular}
    \label{t:params}
\end{table}

All the SSL approaches, but FixMatch, introduce one or more subsidiary terms to the loss. To alleviate their impact at the beginning of the training, these terms are weighted by a lambda $\lambda$ ratio, which ramps up to its maximum value within a warmup length $wl$. The ramp-up strategy is defined in Eq.~(\ref{eq:rampup}) for MT and DCT, and is linear in MM during the first 16k learning iterations.
\begin{equation}
    \label{eq:rampup}
    % \lambda = \lambda_{\text{max}} \times \big( 1 - e^{-5\times (1 - (t / \textrm{wl}))^2} \big)
    % EL - FIX DCT & MT WARMUP STRATEGY (eurasip)
    \lambda = \lambda_{\text{max}} \times e^{-5\times (1 - (t / \textrm{wl}))^2}
\end{equation}

\begin{table*}[th]
    \centering
    \caption{Supervised learning Error Rates (\%) on ESC-10, UBS8K and GSC.}
    \label{t:sl_results}

    % \rowcolors{2}{gray!25}{white}
    
    % \begin{tabular}[0.66\textwidth]{lcccccc}
    \begin{tabular}{lcccccc}
        \toprule
        Dataset                               & \multicolumn{2}{c}{ESC-10}                                                                           & \multicolumn{2}{c}{UBS8K}                                                                            & \multicolumn{2}{c}{GSC}                                                                      \\ 
        \midrule
        % \multicolumn{1}{l|}{Labeled fraction} & 10\%                                    & \multicolumn{1}{c|}{100\%}                                 & 10\%                                   & \multicolumn{1}{c|}{100\%}                                  & 10\%                                                        & 100\%                          \\ 
        
        Labeled fraction & 10\%                      & 100\%       & 10\%                  & 100\%        & 10\%          & 100\% \\
        \midrule
        \multicolumn{1}{l|}{\corrige{CNN models (literature)}} & - & \multicolumn{1}{c|}{\corrige{3.00~\cite{sota_esc10}}} & - & \multicolumn{1}{c|}{\corrige{14.50~\cite{sota_ubs8k}}} & - & \corrige{3.00~\cite{sota_gsc_Vygon_2021}} \\
        % \multicolumn{1}{l|}{CNN-based litterature~\cite{sota_esc10, sota_ubs8k, sota_gsc_Vygon_2021}} & - & \multicolumn{1}{c|}{3.00} & - & \multicolumn{1}{c|}{9.20} & - & 3.00 \\
        \midrule
        \multicolumn{1}{l|}{Supervised}       & 32.00 ± 6.17                            & \multicolumn{1}{c|}{8.00 ± 5.06}                           & 33.80 ± 4.82                           & \multicolumn{1}{c|}{23.29 ± 5.80}                           & 10.01                                                       & 4.94                           \\
        \multicolumn{1}{l|}{+mixup}           & 36.00 ± 5.22                            & \multicolumn{1}{c|}{8.33 ± 4.56}                           & 31.41 ± 5.56                           & \multicolumn{1}{c|}{22.04 ± 5.99}                           & \hphantom{1}8.83                           & 3.86                           \\
        \multicolumn{1}{l|}{+weak}            & \textbf{22.67 ± 3.46} & \multicolumn{1}{c|}{4.67 ± 3.43}                           & 27.08 ± 4.58                           & \multicolumn{1}{c|}{20.09 ± 5.50}                           & \hphantom{1}7.62                           & 3.90                           \\
        \multicolumn{1}{l|}{+weak+mixup}      & 24.67 ± 4.92                            & \multicolumn{1}{c|}{\textbf{4.67 ± 1.39}} & \textbf{23.75 ± 4.73} & \multicolumn{1}{c|}{\textbf{17.96 ± 3.64}} & \textbf{\hphantom{1}6.58} & 3.00 \\
        \multicolumn{1}{l|}{+strong}          & 23.00 ± 5.19                            & \multicolumn{1}{c|}{5.00 ± 2.64}                           & 25.58 ± 4.15                           & \multicolumn{1}{c|}{20.69 ± 4.92}                           & \hphantom{1}7.60                           & 3.27                           \\
        \multicolumn{1}{l|}{+strong+mixup}    & 24.00 ± 8.71                            & \multicolumn{1}{c|}{5.00 ± 4.25}                           & 24.73 ± 4.42                           & \multicolumn{1}{c|}{18.52 ± 4.38}                           & \hphantom{1}6.86                           & \textbf{2.98} \\ \bottomrule
    \end{tabular}
\end{table*}

In MT, the maximum value of $\lambda_{\text{cc}}$ is 1 and $\alpha_\mathrm{ema}$ is set to 0.999. 
In DCT, the maximum values of $\lambda_{\text{cot}}$ and $\lambda_{\text{diff}}$ are 1 and 0.5, respectively. 
In MM the maximum value of $\lambda_u$ is 1.
% EL - EURASIP - Fix confusion in warmup strategy !
\corrige{FM and RMM do not use a ramp up strategy. In FM, the value of $\lambda_u$ is set to 1 and in RMM the values of $\lambda_u$, $\lambda_{u_1}$ and $\lambda_r$ are set to 1.5, 0.5 and 0.5, respectively.}
% In MM and FM the maximum value of $\lambda_u$ is 1.
% and finally, for RMM the maximum value of $\lambda_u$ is 1.5, $\lambda_{u_1}$ and $\lambda_r$ is 0.5.

In MM and RMM, we use two augmentations ($k = 2$), the sharpening temperature $T$ is set to 0.5. 
In FM, we use a threshold $\tau = 0.8$ on ESC-10 and GSC datasets, and $\tau = 0.95$ for UBS8K. 
% EL - EURASIP add history size in hparams
\corrige{In RMM, the number of labels $N$ kept for distribution alignment is set to 128.}
% EL - EURASIP rem duplicate with above
%In RMM, we set the $\lambda_u$ to 1.5 and the $\lambda_{u_1}$ and $\lambda_r$ hyperparameters to 0.5.

For MM, FM and RMM, on ESC-10, the batch size is 60 because ESC-10 is a small dataset of 400 files only. During training, only four folders are used, that is, 320 files. In a 10\% configuration and due to the whole division's restrictions, this represents only 30 supervised files in total. Each mini-batch must contain as many labeled as unlabeled files, hence the batch size of 60. Moreover, because of this small number of files, the training phase only lasts for 2700 iterations, and therefore, warm-up ends prematurely.

For our proposed variants, which include mixup, we kept the same configurations and parameter values. 

% In a 10\% configuration, the labeled files represents 30 examples per batch, because we want to keep an uniform distribution of classes which means we have 3 example per class in the labeled part.

% lc - We use for MM and FM a batch size of 60 for ESC-10 dataset because the labeled part of the dataset contains only 30 examples. The labeled ratio in a batch is 50\% for the three datasets. No learning rate scheduler has been used for these methods.

\section{Results}
\label{sec:results}

We first report the results obtained in a supervised setting, with and without the same data augmentation methods used in the SSL algorithms, including mixup. We compare the error rates obtained by the five SSL methods and then show that adding mixup is almost in all cases beneficial.

\newcommand\score[1]{
    \num[
        round-mode=places, 
        round-precision=2, 
        round-integer-to-decimal=true
    ]{\fpeval{100-#1}}}
    
\newcommand\scorebf[1]{
    \num[
        round-mode=places, 
        round-precision=2, 
        round-integer-to-decimal=true, 
        math-rm=\mathbf
    ]{\fpeval{100-#1}}}

\subsection{Supervised learning}

This section presents the results obtained with supervised learning in different settings while using either 10\% or 100\% of the labeled data available.
MM, RMM and FM use augmentations as their core mechanism. RMM and FM use weak and strong augmentations, while MM uses a combination of weak augmentations and mixup. Therefore, it seems essential for fair comparisons to use the same augmentations in the supervised settings too. 

We trained models without any augmentation (Supervised), using mixup alone (mixup), weak augmentations alone (Weak), a combination of weak augmentations and mixup (Weak+mixup), strong augmentations alone (Strong), and to finish, a combination of strong augmentations with mixup (Strong+mixup). Table~\ref{t:sl_results} presents the results on ESC-10, UBS8K, and GSC. \corrige{In order to give an idea of how our results compare to the literature, we reported three results from the literature, in the ``CNN models (literature)'' row in the table. We chose to report results from works in which the models are primarily based on a CNN architecture, to be fair with the Wide-ResNet we used in our case. There are better results from the recent literature, but that involved large transformer models, sometimes pretrained on AudioSet. For instance, the state-of-the-art result on UBS8K is 10.0\% ER, obtained with a 25-M parameter transformer, pre-trained on AudioSet~\cite{gazneli2022end}.}

\textbf{ESC-10.}
In the 10\% setting, the supervised model reached an ER of 32.00\%. The use of Weak yielded the best performance with 
22.67\% ER, outperforming the supervised model by 9.3 points (29.16\% relative).
In the 100\% setting, the supervised model reached an ER of 8.00\%, and the best ER of 4.67\% was achieved when using Weak+mixup. The gain is 3.33 points (41.62\% relative).

% plus le cas avec nvx results :
%We can also note that weak or strong augmentations alone lead to worse results in a 10\% setting, whereas mixup augmentation did deliver a slight improvement of 1.2\% by reaching 36.44\% error rate. %63.56\% accuracy.

\textbf{UBS8K.}
In a 10\% setting, the supervised model reached 33.80\% ER, and the best supervised result was obtained with Weak+mixup, with a 23.75\% ER. It represents an improvement of 10.05 points, 29.73\% relative improvement.
In the 100\% setting, the same augmentation combination reached an ER of 17.96\%, outperforming the 23.29\% ER from the supervised model by 5.33 points, 22.88\% relative improvement.

\textbf{GSC.}
In a 10\% setting, the supervised model reached 10.01\% ER, and Weak+mixup yielded the best ER of 6.58\% It represents an augmentation of 3.43 points, 34.26\% relative improvement.
In the 100\% setting, the Strong+mixup reached an ER of 2.98\%, outperforming the 4.94\% ER from the supervised model by 1.96 point, 39.68\% relative improvement.

Overall, we observe that in a supervised setting, the combination of mixup with a weak or a strong augmentation is systematically better than using a single augmentation, except in the ESC-10 dataset.

\begin{table*}[tbh]
    \centering
    \caption{Semi-supervised learning Error Rates (\%) on ESC-10, UBS8K and GSC.}
    \label{t:ssl_results}

    \rowcolors{4}{gray!25}{white}
    
    \begin{tabular}{l|cc|cc|cc}
        \toprule
        Dataset                               & \multicolumn{2}{c|}{ESC-10}                         & \multicolumn{2}{c|}{UBS8K}                           & \multicolumn{2}{c}{GSC}               \\
        Labeled fraction                      & 10\%                              & 100\%       & 10\%                  & 100\%        & 10\%          & 100\% \\ 
        \midrule
        Supervised                            & 32.00 ± \hphantom{1}6.17          & 8.00 ± 5.06 & 33.80 ± 4.82          & 23.29 ± 5.80 & 10.01         & 4.94  \\
        Best Supervised                       & 22.67 ± \hphantom{1}3.46          & 4.67 ± 1.39 & 23.75 ± 4.73          & 17.96 ± 3.64 & 6.58          & 2.98  \\ 
        \midrule
        % MT                                    & 28.28 ± \hphantom{1}5.28          & -           & 32.80 ± 4.21          & -            & 8.51          & -     \\
        % MT+mixup                              & 27.81 ± \hphantom{1}2.25          & -           & 32.00 ± 5.80          & -            & 8.50          & -     \\
        % DCT                                   & 25.16 ± \hphantom{1}4.42          & -           & 27.85 ± 4.29          & -            & 6.22          & -     \\
        % DCT+mixup                             & 23.75 ± \hphantom{1}2.36          & -           & 25.77 ± 4.73          & -            & 5.63          & -     \\
        % Update 16/11/2022 : fix MT, MT+mixup, DCT, DCT+mixup GSC results
        MT                                    & 28.28 ± \hphantom{1}5.28          & -           & 32.80 ± 4.21          & -            & 8.92          & -     \\
        MT+mixup                              & 27.81 ± \hphantom{1}2.25          & -           & 32.00 ± 5.80          & -            & 9.32          & -     \\
        DCT                                   & 25.16 ± \hphantom{1}4.42          & -           & 27.85 ± 4.29          & -            & 6.90          & -     \\
        DCT+mixup                             & 23.75 ± \hphantom{1}2.36          & -           & 25.77 ± 4.73          & -            & 5.94          & -     \\
        MM-mixup                              & 17.33 ± \hphantom{1}3.84          & -           & 20.42 ± 4.88          & -            & 4.49          & -     \\
        MM                                    & 15.33 ± \hphantom{1}5.58          & -           & \textbf{18.02 ± 4.00} & -            & \textbf{3.25} & -     \\
        % MM                                    & 15.33 ± \hphantom{1}5.58          & -           & \textbf{18.02 ± 4.00} & -            & 3.25 & -     \\
        % RMM-mixup                             & 35.33 ± 11.93                     & -           & 38.50 ± 5.18          & -            & 4.99          & -     \\
        % RMM                                   & 19.17 ± \hphantom{1}6.52          & -           & 26.02 ± 6.90          & -            & 3.78          & -     \\
        RMM-mixup                             & \corrige{32.50 ± 11.71}                     & -           & \corrige{38.23 ± 6.15}          & -            & \corrige{5.15}          & - \\
        RMM                                   & \corrige{\textbf{12.00 ± \hphantom{1}5.55}} & -           & \corrige{28.41 ± 6.54}          & -            & \corrige{3.54}          & - \\
        % FM                                    & \textbf{13.33 ± \hphantom{1}2.89} & -           & 21.44 ± 4.16          & -            & 4.44          & -     \\
        FM                                    & 13.33 ± \hphantom{1}2.89          & -           & 21.44 ± 4.16          & -            & 4.44          & -     \\
        FM+mixup                              & 14.67 ± \hphantom{1}7.21          & -           & 18.27 ± 3.80          & -            & 3.31          & -     \\
        \bottomrule
    \end{tabular}
\end{table*}

\subsection{Semi-supervised learning}

We report in Table~\ref{t:ssl_results} the results of the SSL methods. For MM and RMM, mixup is already used in the original methods, thus, we compare MM to MM without mixup (MM-mixup) and RMM to RMM without mixup (RMM-mixup). For the three other methods, we denote for instance FM+mixup the FM algorithm augmented with mixup.

In all the three datasets, the five SSL methods brought ER decreases compared to the 10\% supervised learning setup, when no augmentation is performed. Only MM, RMM, and FM performed better than the best supervised training result, that used the weak augmentations. Furthermore, they also significantly outperformed MT and DCT in all but one cases (DCT better than RMM on UBS8K), showing that using single-model SSL methods is more efficient than two-model-based methods, at least on these three datasets and among the five methods that were compared.

\corrige{For ESC-10, in the 10\% setting, the lowest ER was achieved by RMM with a 12.00\% value, compared to a 22.67\% for a weakly augmented supervised training. It represents a 10.67 points improvement, 47.1\% relative. The difference with a fully supervised training using weak augmentations reaching a 4.67\% ER is still notable with a 7.33 points difference.} 

On UBS8K, the best ER was achieved using MM with an 18.02\% ER, very closely followed by FM+mixup with 18.27\%. The difference with the best supervised training Weak+mixup, reaching 23.75\%, represents a difference of 5.73 points (24.13\% relative). The performance of MM is also very close to the best fully supervised training Weak+mixup, which reached a 17.96\% ER. The difference is only 0.06 points. Similarly to ESC-10, if MT and DCT outperformed the supervised training methods, they performed worse than supervised learning with augmentation. \corrige{UBS8K is the only dataset for which RMM performed worse than DCT.}

\corrige{The GSC dataset results confirm the previous observations. The MM method is the best method with an ER of 3.25\%, representing a relative gain of 6.76 (67.53\%) or 3.33 points (50.61\%) compared to supervised training without and with Weak+mixup augmentations, respectively. RMM and FM+mixup obtained results very similar to MM: 3.54\% and 3.31\% ER, respectively.}

\subsection{Impact of mixup}

\corrige{Given that the best SSL methods so far were MM and RMM, and that mixup is used in these approaches, we decided to try to add mixup to MT, DCT, and FM, in different ways for each method as explained in Section~\ref{subsec:addingmixup}.} In~\cite{sohn2020fixmatch}, Appendix D.2, mixup on the entries (not on the labels) was added to FM, removing all the other image augmentations. In this setting, FM was shown to reach an accuracy very close to that of MM on CIFAR-10.

In Table~\ref{t:ssl_results}, we reported the results when adding mixup to MT, DCT and FM, (MT+mixup, DCT+mix\-up, FM+mixup). We also give the ER when removing mixup from MM and RMM, in the row named MM-mixup amd RMM-mixup.

\corrige{As a first comment, MM-mixup and RMM-mixup are always worse than with mixup. For instance, with MM on USB8K, ER increased from 18.02\% to 20.42\%. This is particularly visible with RMM on ESC-10 and UBS8K. Moreover, adding mixup to the other SSL methods brought performance improvements on all the datasets tested. The only counter-example observed is FM on ESC-10, which went from 13.33\% to 14.67\% ER. The standard deviation value also increased significantly from 2.89\% to 7.21\%.

Similarly, FM on UBS8K went from 21.44\% ER without mixup to 18.24\% with mixup. On GSC, RMM presented the largest gap between 5.15\% and 3.54\% ER without and with mixup, respectively.}

% The improvements brought by the addition of mixup did not allow MT and DCT to perform better than augmented supervised training in a 10\% configuration.

It is also important to note that using mixup allowed to get ER values very close to the ones obtained with fully (100\% setting) supervised training using augmentations, on UBS8K and GSC. \corrige{This is observable with MM, RMM, and FM+mixup. For instance, compared to Weak+mixup 100\% supervised, MM has only 0.06 point difference on UBS8K, and 0.27 point difference on GSC.}

When we look at our supervised training performance, we can observe that an improvement does not systematically follow the use of weak or strong augmentations. However, when combined with mixup, ER is frequently improved. This can be partly explained by the fact that audio augmentations are often difficult to choose and that their impact is often dependent on the dataset and the task at hand~\cite{lu2019semi}. With this in mind, mixup seems to be beneficial regardless of the dataset used.

\subsection{Training time}

\begin{figure}[ht]
    % EL - UPDATE DURATIONS SCHEMA with RMM without self-loss (eurasip)
    % \includegraphics[width=0.9\linewidth]{Normalized_durations_without_mixup_v3.png}
    % \includegraphics[width=0.9\linewidth]{img/Durations_v4.png}
    \includegraphics[width=0.9\linewidth]{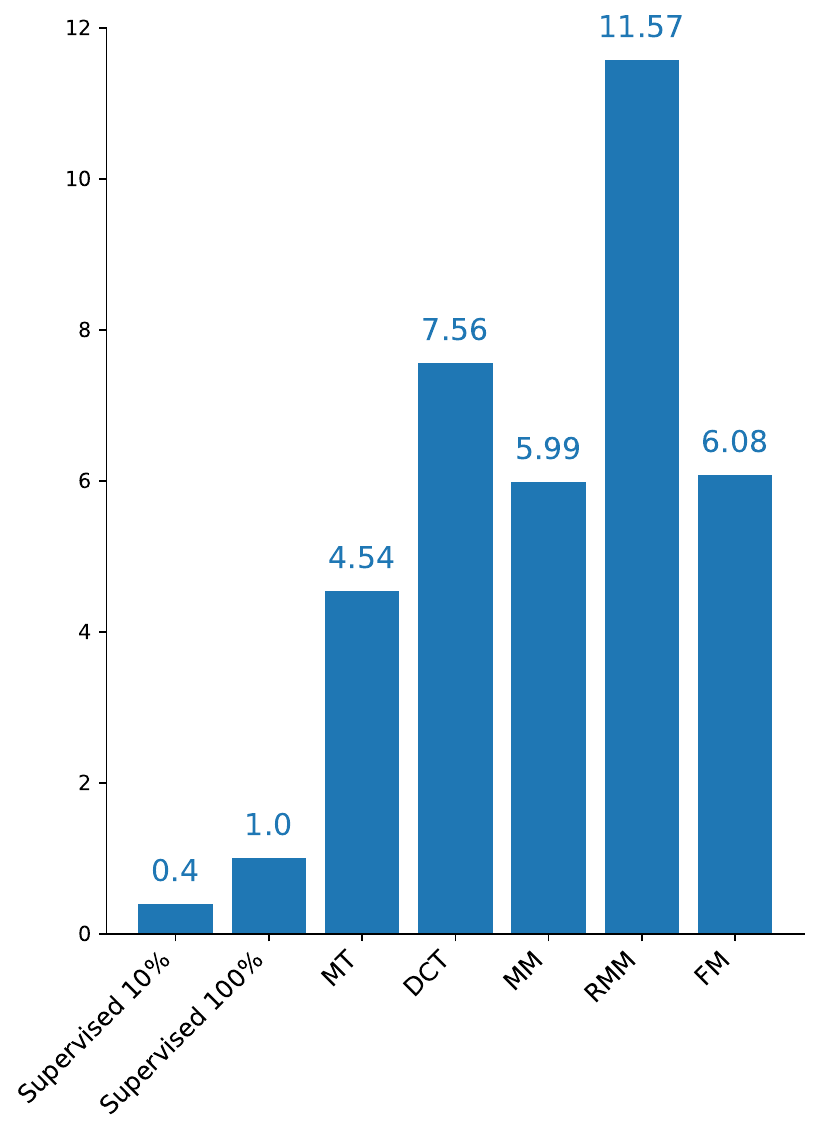}
    \centering
    \caption{Normalized mean training duration for all methods without mixup.}
    \label{fig:durations_means}
\end{figure}

The normalized training duration means for all the five methods are shown in Fig.~\ref{fig:durations_means}. The values were computed on the three datasets using the following equation:
\begin{equation}
    \text{mean} = \frac{d}{N_f \cdot N_e \cdot \text{bs}}
\end{equation}

Where $d$ is the total duration, $N_f$ the number of folds in the dataset, $N_e$ the number of epochs, and $\text{bs}$ the batch size used in each method. We compute the three means for each dataset then we report the average of the three values. Finally, we use the \textit{supervised 100\%} execution time as the reference (training duration of one). We also assessed the impact of adding mixup, but it had a negligible impact of about 0.5\%. % For this reason, we only show the mixup duration of each method.

Among the SSL approaches, the fastest one is MT, which has a training time 4.5 times longer that the fully supervised training. \corrige{Then, FM and MM follow with are 6 times longer. DCT, with its high complexity and use of adversarial data, took up to 7.6 times longer, and finally the longest of all is RMM, 11.6 times longer, due to the large number of augmentations involved.}

% lc - Within the SSL approaches, MT is the fastest one on average, and it takes about 4.5 more times to train than Supervised 100\%. Follow MM and FM, with a factor 6 compared to Supervised 100\%. DCT is the second more demanding approach on average, probably due to the generation of adversarial examples. Finally, RMM is the slowest method probably due to all the processes involved during a training step (flips, 4 different losses components and 2 types of augmentations).

\section{Discussion}

\corrige{\subsection*{Why are MM, RMM and FM better than MT and DCT?} This question remains open.} Several key components may explain this gap in performance. First, data augmentation is extensively used in these methods (weak and strong ones), both on the labeled data and on the unlabeled subset to satisfy the consistency criterion of SSL. No data augmentation is used in the  DCT nor MT basic methods, except the addition of noise in MT, on the unlabeled subset at the input of the teacher model. \corrige{Nevertheless, when mixup was added to MT, no significant gain was observed. Thus, other augmentations should be explored for MT.} Second, MM, RMM and FM use pseudo-labeling, with either explicit entropy minimization (sharpening in MM and RMM) or threshold-based selection (confidence masking in FM). In DCT and MT, no entropy minimization is used, the predictions on the unlabeled part of the data are used as is for a consistency criterion between the two collaborating networks. 

\corrige{\subsection*{Which augmentations?} We used three augmentations (besides mixup): Occlusion, CutOut and Speed perturbation. An advantage of those is that they are task-agnostic. We tuned their hyperparameters once on GSC, and then, we used them on ESC-10 and UBS8K as is, bringing performance improvements. Exploring more audio-specific augmentations is an avenue still to be explored. For instance, we did not try pitch shifting nor dynamic range compression~\cite{Salamon_2017}. Those would need careful parameter tuning depending on the audio event types and on the dataset involved in the experiments.

Finally, Occlusion and CutOut could be replaced by SpecAugment \cite{park2019specaugment}, originally proposed in automatic speech recognition and very often used nowadays in audio processing tasks, such as audio tagging. There is two small differences, though, in using SpecAugment, since it drops out one or several vertical and horizontal stripes from the spectrograms, while CutOut drops out a single rectangle of random shape. Another difference is that we applied randomly either Occlusion or CutOut, but not a combination of the two. To evaluate the effect of SpecAugment, we ran supervised learning experiments on GSC, using Speed Perturbation and mixup, and SpecAugment instead of Occlusion and CutOut, in the 100\% of the labeled training data setting. We tested several configurations for SpecAugment. Our best setting was zero, one or two frequency stripes of width between 0 and 7 bins, and zero or one stripe of width also between 0 and 7 bins in time. This setting led to a 2.51\% ER, which is better than the 2.98\% value of our best supervised baseline method. This confirms experimentally that SpecAugment could replace Occlusion and CutOut, as a combination of the two. We did not rerun all the SSL experiments with SpecAugment, but one might expect slightly better results than those obtained with Occlusion and CutOut.}

% EL - EURASIP
% \corrige{MT and DCT, which use two models, were significantly outperformed by the mono-model  approaches, MM, RMM and FM, except RMM on USB8K.} Several key components of these last three algorithms may explain this gap in performance. \corrige{Data augmentation is extensively used in these methods to satisfy the consistency criterion of SSL. DCT does not use augmentation and MT only applies additive noise to the teacher's model input.} MM, RMM and FM use pseudo-labeling, with either explicit entropy minimization (sharpening in MM and RMM) or a threshold-based selection (confidence masking in FM). In DCT and MT, no pseudo-labeling is used, the predictions on the unlabeled part of the data are used for consistency in the collaborating networks. 

% 3 méthodes donnent des results similaire sauf RMM sur UBS8K
% RMM a atteint les bests perfs sur ESC10
% MM a la meilleur perf sur UBS8K + GSC

% Another interesting result of the present work is the fact that MM and RMM were slightly better than FM in our audio classification experiments. FM was shown to perform better than MM and RMM in the FM original article, on image related tasks. This may be the result of our experimental design choices regarding the so-called weak and strong augmentations. In our work, we use the same augmentations but with stronger rates to distinguish strong ones from weak ones. In the FM article, these augmentations on images are of different nature: image flips-and-shifts for the weak ones and a series of transformations including occlusions for the strong ones.

\section{Conclusions}

In this article, we reported audio classification experiments in a semi-supervised setting on three standard datasets of different sizes and content, the very small-sized ESC-10 with generic audio events, urban noises with UrbanSound8K, and speech with Google Speech Commands. We used only 10\% of the labeled training data samples and the remaining 90\% as unlabeled samples. We adapted and compared five SSL algorithms for this task, two methods that use two neural networks in parallel: Mean Teacher and Deep Co-Training, and the three single-model methods MixMatch, ReMixMatch and FixMatch, that strongly rely on data augmentation.

\corrige{
All the five methods brought significant gains compared to a supervised training setting using 10\% of labeled data. They performed better than supervised learning without augmentation. On UBS8K, MixMatch and FixMatch were very close to fully supervised learning with augmentation (100\% of labeled training data).
On ESC-10, ReMixMatch reached the best Error Rate of 12.00\%. The relative gains were 62\% and 47\%, when compared to a supervised training using 10\% of labeled data, without and with augmentation, respectively.
On UrbanSound8K, MixMatch obtained the best results, reaching 18.02\% Error Rate. Compared to a 10\% supervised training without and with augmentation, the respective relative improvements were 47\% and 24\%.
On Google Speech Commands, MixMatch again reached the best Error Rate of 3.25\%. The relative improvement was 68\% and 51\%, compared to a 10\% supervised training without and with augmentation, respectively.
}
Mixup is an efficient regularization technique that is at the heart of the MixMatch and ReMixMatch algorithms. Its consistent impact in MM and RMM encouraged us to add it to the other SSL approaches. In almost all the experiments, adding mixup brought consistent improvements, which allowed us to get closer to the best supervised learning settings using 100\% of the labeled data available. For instance, adding mixup to FixMatch reduced the error rates on UrbanSound8K from 21.4\% to 18.3\%, and from 4.4\% to 3.3\% on Google Speech Commands, to be compared with 17.9\% and 3.0\% respectively, obtained in the best supervised learning settings. 
% the difference between supervised learning and SSL was 0.31 points (1.70\%) only for UrbanSound8K and Google Speech Commands (10.33\%).

\corrige{
In conclusion, if we were to recommend a method out of the ones tested in our work, we would recommend MixMatch, and FixMatch+mixup also, with very similar performances. Their good results are consistent across the three datasets. The gains brought by these methods is worth their training time, about six times the 100\% supervised setting training time. ReMixMatch obtained the best results on ESC-10, but this method is more demanding in training time.
}

Many questions remain open, though. The fact that MM and RMM were slightly better than FM needs to be further investigated, in particular the use of audio augmentations different in nature for the weak and the strong ones may be a direction to explore. \corrige{MT and DCT do not use augmentations in their original version. It would be interesting, though, to try the weak augmentations used in the holistic methods with them.} We also plan to adapt the SSL methods to multi-label audio tagging, for instance on Audioset~\cite{audioset} or FSD50K~\cite{fonseca2020fsd50k}. In particular, we would have to adapt the sharpen method in MixMatch, and the thresholding operations in FixMatch. Finally, new SSL methods have been very recently proposed and could be added to our list, such as Unsupervised Data Augmentation (UDA)~\cite{xie2020unsupervised}, and the recent Meta Pseudo Labels method~\cite{pham2021meta}.

% would be appropriate datasets which provides a more general sound context due to its considerable size and complexity (over two million files, 527 classes, multi-labels). Other semi-supervised approaches may be added as we go along. We are thinking of UDA, which has been applied to images and text and presented impressive performances. We could also add more straightforward approaches like pseudo-labeling or temporal ensembling.



% \section*{Appendix}
% Text for this section\ldots

%%%%%%%%%%%%%%%%%%%%%%%%%%%%%%%%%%%%%%%%%%%%%%
%%                                          %%
%% Backmatter begins here                   %%
%%                                          %%
%%%%%%%%%%%%%%%%%%%%%%%%%%%%%%%%%%%%%%%%%%%%%%

\begin{backmatter}

\section*{Acknowledgements}%% if any
Not applicable

\section*{Funding}%% if any
This work was partially supported by the French ANR agency within the LUDAU project (ANR-18-CE23-0005-01) and the French "Investing for the Future --- PIA3" AI Interdisciplinary Institute ANITI (Grant agreement ANR-19-PI3A-0004). We used HPC resources from CALMIP (Grant 2020-p20022) and from the Osirim platform.

% \section*{Abbreviations}%% if any
% Text for this section\ldots

\section*{Availability of data and materials}%% if any
The three datasets used in this article are publicly archived datasets:
https://github.com/karolpiczak/ESC-50
https://urbansounddataset.weebly.com/urbansound8k.html
http://download.tensorflow.org/data/speech\_commands\_v0.02.tar.gz

% \section*{Ethics approval and consent to participate}%% if any
% Text for this section\ldots

\section*{Competing interests}
The authors declare that they have no competing interests.

% \section*{Consent for publication}%% if any
% Text for this section\ldots

\section*{Authors' contributions}
LC implemented the DCT and MT methods, wrote their descriptions. EL implemented the holistic methods (MM, RMM and FM) and wrote their descriptions. They both conducted all the experiments. They were both major contributors in writing the manuscript. TP did the conception and design of the work, wrote the related work section, and substantively revised the manuscript. All authors read and approved the manuscript.
% \section*{Authors' information}%% if any
% Text for this section\ldots

%%  After submission of the .TEX file,                     %%
%%  you will be prompted to submit your .BBL file.         %%

% if your bibliography is in bibtex format, use those commands:
\bibliographystyle{bmc-mathphys} % Style BST file (bmc-mathphys, vancouver, spbasic).
\bibliography{bmc_article}      % Bibliography file (usually '*.bib' )

\end{backmatter}
\end{document}